\documentclass[10pt]{iopart}

\usepackage{iopams}  

\expandafter\let\csname equation*\endcsname\relax

\expandafter\let\csname endequation*\endcsname\relax

\usepackage{amsmath}
\usepackage{xcolor} 
\usepackage{comment}
\usepackage{graphicx} 
\usepackage{hyperref}

\renewcommand\[{\begin{equation}}
\renewcommand\]{\end{equation}}

\global\long\def\ket#1{\left|#1\right\rangle }%
\global\long\def\bra#1{\left\langle #1\right|}%
\global\long\def\abs#1{\left|#1\right|}%
\global\long\def\braket#1#2{\left\langle #1|#2\right\rangle }%

\usepackage{color,soul}
\usepackage{tikz}
\usepackage{verbatim}
\usepackage{float}  
\usepackage{graphicx}
\usepackage{dsfont}
\usepackage{color}
\usepackage{mathdots}
\usepackage{mathtools}
\usepackage[english]{babel}
\usepackage{booktabs}
\usetikzlibrary{fadings}
\usetikzlibrary{patterns}
\usetikzlibrary{shadows.blur}
\usetikzlibrary{shapes}
\usetikzlibrary{matrix}
\usepackage{caption}
\usepackage{url}           
\usepackage{bm}
\newtheorem{definition}{Definition}[section]
\usepackage{framed}
\usepackage{braket}
\usepackage{bbold}
\usepackage{graphicx}
\usepackage{dcolumn}
\usepackage{bm}
\usepackage{hyperref}
\usepackage{amsmath}
\usepackage{calrsfs}
\DeclareMathAlphabet{\pazocal}{OMS}{zplm}{m}{n}
\newcommand{\HC}{\pazocal{H}}
\newcommand{\BC}{\pazocal{B}}
\newcommand{\FC}{\pazocal{F}}
\newcommand{\DC}{\pazocal{D}}

\newcommand{\ACAL}{\pazocal{A}}
\newcommand{\NCAL}{\pazocal{N}}

\usepackage{dsfont}
\usepackage{braket}
\usepackage{makecell}
\usepackage{colortbl}

\newtheorem{theorem}{Theorem}[section]

\begin{document}

	\title{Mode Entanglement in Fermionic and Bosonic Harmonium}
	\author{Jan Ole Ernst, Felix Tennie}
 	\address{
		University of Oxford, Department of Physics, Clarendon Laboratory, Parks Road, Oxford OX13PU\\
	}
 \ead{felix.tennie@physics.ox.ac.uk}
	

	
	
	
	\date{\today}


	\begin{abstract}
		Mode entanglement in many-body quantum systems is an active area of research. It provides crucial insight into the suitability of many-body systems for quantum information processing tasks. Local super-selection rules must be taken into account when assessing the amount of physically accessible entanglement. This requires amending well-established entanglement measures by incorporating local parity and local particle number constraints. In this paper, we report on mode entanglement present in the analytically solvable system of $N$-Harmonium. To the knowledge of the authors, this is the first \textit{analytic} study of the physically accessible mode and mode-mode entanglement of an interacting many-body system in a continuous state space. We find that super-selection rules dramatically reduce the amount of physically accessible entanglement, which vanishes entirely in some cases. Our results strongly suggest the need to re-evaluate intra and inter-mode entanglement in other fermionic and bosonic systems.

	\end{abstract}
	

	\maketitle
	
\ioptwocol

	\section{Introduction}
    Entanglement is ``the characteristic trait of quantum physics" \cite{schrodinger_1935} and despite recent advances still remains somewhat elusive. It provides insights into the behaviour of physical systems and forms a cornerstone for the investigation of many-body quantum physics \cite{ent_manybody}. Therefore, describing and quantifying entanglement in many-body systems is necessary for understanding and developing adequate models. Furthermore, entanglement is a key resource for quantum information processing, particularly for the realisation of quantum computers \cite{deutsch_QC}. It is widely conjectured that systems exhibiting only a limited amount of entanglement could be emulated by existing noisy quantum simulation and computation technologies \cite{Preskill2018}.

    Fermionic mode entanglement plays a crucial role in quantum chemistry \cite{Ding2020, Ding2020a}. Moreover, following advances in single electron manipulation \cite{hofer2017}, fermionic mode entanglement offers a potential resource for quantum teleportation schemes and information processing tasks \cite{Galler, Olofsson2020, Debarba2020, ferm_mbqc}. Mode entanglement is also highly relevant in systems of cold bosonic atoms \cite{Gross2011}, and in a continuous variable setting, error correction codes show promise for the realisation of fault-tolerant quantum computing \cite{gkp_states}.

    The assessment and quantification of entanglement in systems of indistinguishable particles, i.e.~fermions and bosons, has been the subject of many controversial debates \cite{BENATTI20201, dalton2016quantum, morris-indist, cunha-mode}. While some states, such as the two-particle state $(|01\rangle \pm |10\rangle)$, appear to contain entanglement in a `first-quantisation' framework, it turns out that this form of entanglement is generally not accessible by physical measurements. 
    
    Recently, Ding et al.~\cite{Ding2020} worked out an approach that consistently generalises well-known entanglement concepts and measures to systems of indistinguishable particles. They then evaluated these new adapted measures in analytically solvable low dimensional fermionic model systems, and provided a numerical study of molecular systems by truncating their infinite dimensional active spaces. 
    
    Despite these numerical results, no \textit{analytic} study on the generalised entanglement measures introduced by Ding et al.~for interacting many-body systems with a continuous state space has been conducted so far. In most cases it is not possible to determine an analytic form of the energy eigenstates of an interacting many-body Hamiltonian. However, the system of Harmonium is a notable exception. Harmonium is a model system of harmonically interacting particles confined by a harmonic trapping potential. To leading order, it resembles the Hamiltonians of systems such as quantum dots \cite{Moshinsky1968, cioslowski2006, laguna2011, nagy2011, pipek2009, bouvrie, johnson91, teichmann2013}). Previous work has considered \textit{particle} entanglement of the Harmonium ground state \cite{Benavides-Riveros2014}. In this paper we investigate the amount of \textit{physically accessible} entanglement in Harmonium. In particular, we evaluate mode and mode-mode entanglement of the Harmonium ground state. Remarkably, the analytic form of the ground state permits an analytic evaluation of the generalised entanglement measures. We find that the amount of physically accessible entanglement is significantly smaller than the apparent amount of so-called `fluffy-bunny' entanglement \cite{fluffy_bunny}.
    
    This paper is organised as follows: in section II we introduce correlation and entanglement measures and review the concepts of particle and mode-reduced density operators. We also describe the model system of Harmonium and discuss the role and status of (local) super-selection rules. In section III we present our results of mode and mode-mode entanglement in fermionic and bosonic Harmonium. In section IV we provide a discussion and outlook.

	\section{Background}
	\subsection{Correlation and Entanglement Measures}
\label{sec:measures}
The state of any quantum system can be described by a positive semi-definite density operator $\rho = \sum_{i} p_{i}\ket{\psi_{i}} \bra{\psi_{i}}$ with unit trace $\sum_{i}p_{i}=1$ \cite{nielsen00}. Physical observables are associated with operators $\hat{O}$ which form an algebra $\ACAL(\HC)$. The expectation value of any observable $\hat{O}$ with respect to $\rho$ reads:
\begin{equation}
	\Braket{\hat{O}}_{\rho} = tr(\rho \hat{O}).
\end{equation}
Provided that the total Hilbert space can be expressed as a product of two Hilbert spaces $\HC = \HC_{A}\otimes \HC_{B}$, and that the set of physical observables form an algebra which correspondingly decomposes as $\ACAL(\HC) = \ACAL(\HC_{A}) \otimes \ACAL(\HC_{B})$,
the reduced density operator $\rho_{A}=tr_{B}(\rho)$ of a particular subsystem $A$ is uniquely determined by the following condition on the observables $ \hat{O}_{A}$  \cite{nielsen00}:
\begin{equation} \label{eqn:obs}
\begin{aligned}
	\Braket{ \hat{O}_{A}}_{\rho_{A}} = \Braket{ \hat{O}_{A} \otimes \mathbb{1}_{B} }_{\rho} , \forall  \hat{O}_{A} \in \ACAL(\HC_{A}).
\end{aligned}
\end{equation}

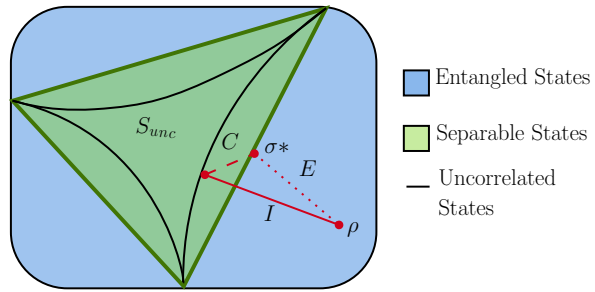
\begin{figure}[h]

\tikzset{every picture/.style={line width=0.75pt}} 

\begin{tikzpicture}[x=0.75pt,y=0.75pt,yscale=-0.5,xscale=0.5]

\draw  [fill={rgb, 255:red, 74; green, 144; blue, 226 }  ,fill opacity=0.53 ] (56.33,69) .. controls (56.33,37.61) and (81.78,12.17) .. (113.17,12.17) -- (368.17,12.17) .. controls (399.55,12.17) and (425,37.61) .. (425,69) -- (425,239.5) .. controls (425,270.89) and (399.55,296.33) .. (368.17,296.33) -- (113.17,296.33) .. controls (81.78,296.33) and (56.33,270.89) .. (56.33,239.5) -- cycle ;
\draw  [color={rgb, 255:red, 65; green, 117; blue, 5 }  ,draw opacity=1 ][fill={rgb, 255:red, 126; green, 211; blue, 33 }  ,fill opacity=0.41 ][line width=1.5]  (374.29,13.61) -- (230.66,294.24) -- (58,107) -- cycle ;
\draw    (58,107) .. controls (58.24,106.28) and (88.78,118.74) .. (140.7,115.54) .. controls (192.62,112.33) and (219.41,104.3) .. (270.36,79.71) .. controls (321.32,55.11) and (336.98,44.04) .. (374.29,13.61) ;
\draw    (230,290.33) .. controls (229.34,107.1) and (373.33,14.33) .. (374.29,13.61) ;
\draw    (58,107) .. controls (210.67,141.67) and (233.33,293.33) .. (230,290.33) ;
\draw [color={rgb, 255:red, 208; green, 2; blue, 27 }  ,draw opacity=1 ][fill={rgb, 255:red, 0; green, 0; blue, 0 }  ,fill opacity=1 ]   (252,181.67) -- (387.33,232.33) ;
\draw [shift={(387.33,232.33)}, rotate = 20.53] [color={rgb, 255:red, 208; green, 2; blue, 27 }  ,draw opacity=1 ][fill={rgb, 255:red, 208; green, 2; blue, 27 }  ,fill opacity=1 ][line width=0.75]      (0, 0) circle [x radius= 3.35, y radius= 3.35]   ;
\draw [shift={(252,181.67)}, rotate = 20.53] [color={rgb, 255:red, 208; green, 2; blue, 27 }  ,draw opacity=1 ][fill={rgb, 255:red, 208; green, 2; blue, 27 }  ,fill opacity=1 ][line width=0.75]      (0, 0) circle [x radius= 3.35, y radius= 3.35]   ;
\draw [color={rgb, 255:red, 208; green, 2; blue, 27 }  ,draw opacity=1 ] [dash pattern={on 0.84pt off 2.51pt}]  (302,160.33) -- (387.33,232.33) ;
\draw [shift={(302,160.33)}, rotate = 40.16] [color={rgb, 255:red, 208; green, 2; blue, 27 }  ,draw opacity=1 ][fill={rgb, 255:red, 208; green, 2; blue, 27 }  ,fill opacity=1 ][line width=0.75]      (0, 0) circle [x radius= 3.35, y radius= 3.35]   ;
\draw [color={rgb, 255:red, 208; green, 2; blue, 27 }  ,draw opacity=1 ] [dash pattern={on 4.5pt off 4.5pt}]  (252,181.67) -- (302,160.33) ;
\draw  [fill={rgb, 255:red, 74; green, 144; blue, 226 }  ,fill opacity=0.65 ] (452.67,76.33) -- (478,76.33) -- (478,101.67) -- (452.67,101.67) -- cycle ;
\draw  [fill={rgb, 255:red, 126; green, 211; blue, 33 }  ,fill opacity=0.43 ] (452.67,132.33) -- (478,132.33) -- (478,157.67) -- (452.67,157.67) -- cycle ;
\draw    (455.33,193.67) -- (478.67,193.67) ;

\draw (394,227.67) node [anchor=north west][inner sep=0.75pt]  [font=\LARGE,xscale=0.5,yscale=0.5]  {$\rho $};
\draw (310.67,147.67) node [anchor=north west][inner sep=0.75pt]  [font=\LARGE,xscale=0.5,yscale=0.5]  {$\sigma *$};
\draw (345.33,165.67) node [anchor=north west][inner sep=0.75pt]  [font=\LARGE,xscale=0.5,yscale=0.5]  {$E$};
\draw (310,210.67) node [anchor=north west][inner sep=0.75pt]  [font=\LARGE,xscale=0.5,yscale=0.5]  {$I$};
\draw (267.33,138.33) node [anchor=north west][inner sep=0.75pt]  [font=\LARGE,xscale=0.5,yscale=0.5]  {$C$};
\draw (180.67,122) node [anchor=north west][inner sep=0.75pt]  [font=\LARGE,xscale=0.5,yscale=0.5]  {${{S_{u}}_{n}}_{c}$};
\draw (482.67,72.67) node [anchor=north west][inner sep=0.75pt]  [font=\LARGE,xscale=0.5,yscale=0.5] [align=left] {Entangled States};
\draw (484,129.33) node [anchor=north west][inner sep=0.75pt]  [font=\LARGE,xscale=0.5,yscale=0.5] [align=left] {Separable States};
\draw (486,175) node [anchor=north west][inner sep=0.75pt]  [font=\LARGE,xscale=0.5,yscale=0.5] [align=left] {Uncorrelated\\States};

\end{tikzpicture}
\caption{Visual representation of sets of quantum states and correlation measures  \cite{Ding2020}.}
\label{fig:correlation}
\end{figure}

Within the set of quantum states, one distinguishes uncorrelated states, correlated states, separable states and entangled states. The notion of correlation and entanglement crucially relies on the product structure of the algebra of observables. For an uncorrelated state, all expectation values factorise into expectation values of local measurements on the subsystems:

\begin{definition}[Uncorrelated states]
	A state represented by density operator $\rho$ is uncorrelated iff:
	$\Braket{\hat{O}_{A} \otimes \hat{O}_{B}}_{\rho} = \Braket{\hat{O}_{A}}_{\rho_{A}}\Braket{\hat{O}_{B}}_{\rho_{B}}, \hat{O}_{A/B} \in \ACAL(\HC_{A/B})$. The set of uncorrelated states is denoted by ${S_{unc}}$. A state is correlated iff its density operator $\rho \not\in S_{unc}$.
\end{definition}
The notion of separable states allows one to distinguish classical correlations and quantum entanglement. The set of all classical mixtures of uncorrelated states forms the set of separable states. States which are not separable are called entangled. More formally:
\begin{definition}[Separable and entangled states]
A bipartite state $\rho$ is separable iff it can be represented as:
$\rho = \sum_{i} p_{i} (\rho_{A}^{i} \otimes \rho_{B}^{i})$ for a probability distribution ${p_{i}}$ and density operators $\rho_{A/B} \in \HC_{A/B}$. Denote the set of these states by ${S_{sep}}$. A state $\rho$ is called entangled iff $\rho \not\in S_{sep}$. 
\end{definition}
Pure, uncorrelated states are necessarily separable. Geometrically, the set of separable states is the convex hull of the set ${S_{unc}}$ as illustrated in Fig.~\ref{fig:correlation}.\footnote{The separability of a state can be assessed by a number of criteria. In this paper, we shall use the PPT criterion \cite{ppt_separability} which is a necessary and sufficient condition for separability in two-dimensional subsystems and provides a lower bound for systems in larger state spaces.}

Classical correlations are quantified by the Shannon entropy. Its quantum analogue is the von Neumann entropy:
\begin{equation}
	S(\rho) := -tr[\rho \ln (\rho)] = - \sum_{i}\lambda_{i}\ln(\lambda_{i}),
\end{equation}
where $\lambda_{i}$ are the eigenvalues of $\rho$ \cite{nielsen00}. The quantum relative entropy is a measure which captures our ability to distinguish two quantum states $\rho$ and $\sigma$:
\begin{equation}
	S(\rho \| \sigma) := tr(\rho \; ln(\rho) -\rho \; ln(\sigma)).
\end{equation}
 The quantum mutual information 
 $I(\rho)$ is a widely accepted measure of the total correlation \cite{Henderson2001, Ding2020} between two subsystems $A$ and $B$:
\begin{equation} \label{eqn:mutinf}
	I(\rho) \equiv I(A:B) := S(\rho_{A}) + S(\rho_{B}) - S(\rho) = S(\rho \| \rho_{A} \otimes \rho_{B}).
\end{equation}

Geometrically, the quantum mutual information describes the distance to the closest uncorrelated state \cite{Ding2020} $I(\rho) = min_{\sigma \in S_{unc}} S(\rho \| \sigma)$.
 We consider the relative entropy of entanglement, $E$ \cite{Piani2009} as our measure of entanglement. $E$ is defined as the quantum relative entropy of the state $\rho$ and the closest separable state $ \sigma* \in {S_{sep}}$:
\begin{equation} \label{eqn:relent} 
	E(\rho) = \min_{\sigma \in S_{sep}} S(\rho \| \sigma) = S(\rho \| \sigma*),
\end{equation}
 $E$ is strictly positive if and only if $\rho$ is non-separable, i.e. entangled \cite{Piani2009}. Geometrically, the relative entropy of entanglement describes the distance to the closest separable state $\sigma *$, cf.\ the dotted red line in Fig. ~\ref{fig:correlation}. The relative entropy of entanglement for a bipartite system in a pure state simplifies to $E(\rho) = S(\rho_{A/B})$. 
 
The classical correlation of a state $\rho$ is defined as the quantum relative entropy of the separable state $\sigma*$ and the closest uncorrelated state $\rho_A \otimes \rho_B$ \cite{Henderson2001}:
\begin{equation}\label{eqn:clcor}
	C(\rho) = S(\sigma* \| \rho_{A} \otimes \rho_{B}).
\end{equation}
The classical correlation can be visualised by the geometrical distance between $\sigma*$ and the closest uncorrelated state.
In general, the amount of entanglement and classical correlation does not sum up to the total correlation $(I \neq E + C)$ as mixed quantum states give rise to correlation described by the quantum discord \cite{Adesso2016}.
\subsection{Tracing out particles and modes}
	In this section, we outline the concepts of tracing out particles and modes; further details can be found in Ref.~\cite{Amosov2017a}. It should be noted that mode reduced density operators fundamentally differ from particle reduced density operators. 
 
	For a state of fixed particle number, such as the $N$-Harmonium ground state, the $p$-particle reduced density operator reads:
	\begin{equation}\label{eq:part_red_dens_operator}
	\begin{aligned}
	    \rho^{(p)} :=& \int_{\mathcal{R}} \psi(x_1,\dots, x_p, x_{p+1}, \dots, x_N)  \\  \cdot &\psi^{*}(x_1', \dots, x_p', x_{p+1}, \dots, x_N) dx_{p+1} \dots dx_N .
	    	\end{aligned}
	\end{equation}
Particle reduced density operators are well defined in systems of both distinguishable and indistinguishable particles. They correspond to observables of fixed particle number $p$.

Observations on a reduced set of modes give rise to the notion of mode-reduced density operators. In systems of many indistinguishable particles, the total Hilbert space of (anti-)symmetrised states does not have an apparent product structure. Let $\lbrace {\ket{\phi_{i}^{1}}}_{i=1}^{\DC} \rbrace$ form a basis of modes of the single-particle Hilbert space $\HC^{(1)}$. Observations confined to a subset of modes $\lbrace{\ket{\phi_{i}^{(1)}}}_{i=1}^{\DC^{A}}\rbrace$ are related to a decomposition of the single-particle Hilbert space into an orthogonal sum:
    \begin{equation}
  	\HC^{(1)} = \HC^{A} \oplus \HC^{B} = sp\lbrace {\ket{\phi_{i}^{(1)}}}_{i=1}^{\DC^{A}} \rbrace\; \oplus \; sp\lbrace {\ket{\phi_{i}^{(1)}}}_{i=\DC^{A}+1}^{\DC} \rbrace .
    \end{equation}

In the following, we shall specifically outline the steps to trace out modes in systems of fermions and bosons. Creation and annihilation operators on the subspaces $\HC^{A}$ and $\HC^{B}$ will allow us to construct a product Hilbert space that is isomorphic to the total Hilbert space.

  		\subsubsection{Fermionic Mode-Reduced Density Operators}

    Fermionic annihilation and creation operators satisfy canonical anti-commutation rules \cite{Jordan1928}:
  
  		\begin{equation}
  		  \label{eq:ferm_creat}
  		\begin{aligned}
  		        f_i f_j^{\dag} + f_j^{\dag}f_i =& \delta_{ij} \mathds{1}, \\
  		        f_i f_j + f_j f_i=& f_i^{\dag}f_j^{\dag}+f_j^{\dag}f_i^{\dag}=0.
  		\end{aligned}
  		\end{equation}
    They can be used to define occupation number states, which form a basis of the Fock space $\FC$ associated with $\HC^{(1)} = sp \lbrace {\ket{\phi_{i}^{(1)}}}_{i=1}^{\DC} \rbrace $,
    
  	\begin{equation}\label{eq:creat}
  	\begin{aligned}
  	\ket{n_1, \dots, n_{\DC}} := (f_1^{\dag})^{\left( n_1 \right)} \dots (f_{\DC}^{\dag})^{\left(n_\DC \right)}\ket{\Omega}.
  	\end{aligned}
  	\end{equation} 
Here, $\ket{\Omega}$ denotes the vacuum state and $n_{i} = \lbrace 0,1 \rbrace$. Fermionic creation and annihilation operators associated with the modes of $H^{A}$ and $H^{B}$ give rise to Fock spaces $\FC_A$ and $\FC_B$. A natural isomorphism  between the product of these Fock spaces and the total Fock space 
  \begin{equation}
  	\FC \cong \FC_A \otimes \FC_B
  \end{equation} 
can be constructed by mapping number states as follows:
  	\begin{equation} \label{eqn:configstate}
  		\ket{n_1, \dots, n_{\DC}} \mapsto \underbrace{\ket{n_1, \dots, n_{\DC_A}}}_{\FC_{A}} \otimes \underbrace{\ket{n_{\DC_{A+1}}, \dots, n_{\DC}}}_{\FC_{B}}.
  		\end{equation}
Note that for the isomorphism to be well-defined, a strict order of the mode labels must be adhered to. This becomes obvious when swapping two creation operators which introduces an additional overall minus sign, cf.~Eq.(\ref{eq:ferm_creat}). 

Given the product structure of the isomorphic Hilbert space $\FC_A \otimes \FC_B$ it is possible to define reduced density operators for specific subsets of modes. In this paper, we consider mode reduced and mode-mode reduced density operators. Using the canonical anti-commutation relations \eqref{eq:ferm_creat}, matrix representations of the reduced density operators can be formed out of expectation values. 

The mode-reduced density matrix of mode $\ket{\phi_m}$ in the local basis $\{ \ket{\Omega}, f^{\dagger}_m \ket{\Omega} \}$ reads \cite{Amosov2017a}:

	\begin{equation}
	\label{eq:1moderdm}
  		       \rho_{\{m\}}=\begin{pmatrix}
  		        \left\langle f_m f_m^{\dag} \right\rangle  & \left\langle f_m^{\dag} \right\rangle \\
  		           \left\langle f_m \right\rangle & \left\langle f_m^{\dag}f_m \right\rangle 
  		    \end{pmatrix}.
  		\end{equation}

  	The mode-mode reduced density matrix of modes $\ket{\phi_{i}}$ and $ \ket{\phi_{j}}$ in the local basis $\{ \ket{\Omega}, f^{\dagger}_i \ket{\Omega}, f^{\dagger}_j \ket{\Omega}, f^{\dagger}_i f^{\dagger}_j\ket{\Omega} \}$ is given by:
  		\begin{equation} 
  		\label{eq:2moderdm}
  		\begin{aligned}
  		    &\; \; \; \rho_{\{i,j\}}= \\
  		    &\begin{pmatrix}
  		           \left\langle f_{i}f_{i}^{\dag} f_{j}f_{j}^{\dag} \right\rangle  & \left\langle f_{i} f_{i}^{\dag} f_{j}^{\dag} \right\rangle  & \left\langle f_{i}^{\dag}f_{j} f_{j}^{\dag} \right\rangle  & \left\langle f_{i}^{\dag} f_{j}^{\dag} \right\rangle\\
  		           \left\langle f_{i} f_{i}^{\dag}f_{j} \right\rangle & \left\langle f_{i}f_{i}^{\dag}f_{j}^{\dag}f_{j} \right\rangle & \left\langle f_{i}^{\dag}f_{j} \right\rangle & \left\langle f_{i}^{\dag}f_{j}^{\dag}f_{j} \right\rangle \\
  		           \left\langle f_if_{j} f_{j}^{\dag} \right\rangle & \left\langle f_{j}^{\dag}f_{i} \right\rangle & \left\langle f_{i}^{\dag}f_{i} f_{j}f_{j}^{\dag} \right\rangle & \left\langle f_{i}^{\dag}f_{j}^{\dag}f_{i} \right\rangle \\
  		            \left\langle f_{j}f_{i} \right\rangle & \left\langle f_{i}f_{j}^{\dag}f_{j} \right\rangle & \left\langle f_{i}^{\dag}f_{j}f_{i} \right\rangle & \left\langle f_{i}^{\dag}f_{i}f_{j}^{\dag}f_{j} \right\rangle
  		    \end{pmatrix}.
  		     \end{aligned}
  		\end{equation}

    It should be noted that, in general, fermionic modes comprise spatial and spin degrees of freedom. Therefore, for a spin-$1/2$ particle, the reduced density operator of a spatial mode $\phi$ is given by Eq.~\eqref{eq:2moderdm} with $i\equiv\{\phi,\uparrow\}$ and $j\equiv\{\phi,\downarrow\}$. The reduced density operator of two spatial modes becomes a $16\times 16$ matrix. In this paper, we explicitly compute both types of reduced density operators for the ground state of Harmonium. Further details will be presented in Sec.~\ref{sec:results} and the evaluation of the matrix elements for the Harmonium ground state is outlined in \ref{sec:expval_int}.

  		\subsubsection{Bosonic Mode-Reduced Density Operators}
  		\label{sec:bos-modered}

    Reduced density operators for subsets of modes can also be defined in bosonic systems. Bosons satisfy the well-known canonical commutation relations \cite{Jordan1928}:
  	    \begin{equation}
  		\begin{aligned}
  		    		b_i b_j^{\dag} - b_j^{\dag}b_i =& \delta_{ij} \mathds{1}, \\
  		    		 b_i b_j - b_j b_i=& b_i^{\dag}b_j^{\dag}-b_j^{\dag}b_i^{\dag}=0.
  		\end{aligned}
  		\end{equation}
  		These can be used to define an (orthonormal) number-state basis of the bosonic Fock-space $\BC$:
  		\begin{equation}
  		\ket{n_1, \dots, n_{\DC}}_{b} = \frac{(b_i^{\dag})^{\left( n_1 \right)}}{\sqrt{n_1 !}} \dots \frac{(b_{\DC}^{\dag})^{\left(n_\DC \right)}}{\sqrt{n_\DC !}}\ket{\Omega},
  		\end{equation}
  		where $n_{i}= \lbrace 0, \cdots, N \rbrace$ .

  	Since the occupation number of bosonic modes is not constrained by the Pauli exclusion principle, the general expression of mode-reduced density operators in terms of expectation values of creation and annhilation operators is more complex. 
   
   For instance, the matrix elements of the mode-reduced density operator of mode $\ket{\phi_m}$ in the local basis 
        $   \{ \ket{\Omega}, b_m^{\dagger}\ket{\Omega}, \frac{(b_m^{\dagger})^2}{\sqrt{2 !}}\ket{\Omega}, \ldots\} $ read:
  		\begin{equation}
  		\label{eq:bos_exp_mode}
        \begin{aligned}
        \left( \rho_{\{ m \}} \right)_{(k,l)} =  
        \frac{1}{\sqrt{k! \;  l!}}  \Bigg\langle (b^{\dagger}_m)^{k} \left(\prod_{i=1}^{N_{max}} (1- b^{\dagger}_m b_m/i ) \right)(b_m)^{l} \Bigg\rangle
        \end{aligned}
  		\end{equation}
    Here, $N_{max}$ represents the highest particle number sector in which the support of the global bosonic state does not vanish. In particular, for pure states of fixed particle number such as the $N$-Harmonium ground state, $N_{max}=N$. Also, note that for finite $N$, $\rho_{\{ m \}}$ is a $(N+1) \times (N+1)$ matrix.

    Similarly, bosonic mode-mode reduced density operators $\rho_{\{ i,j \}}$ can be expressed as expectation values of creation and annihilation operators. However, the size of $\rho_{\{ i,j \}}$ scales quadratically with the particle number $N$.

    Unlike particle-reduced density operators, mode-reduced density operators can be mixed states of various particle numbers, as evident by Eqs.~\eqref{eq:1moderdm}, \eqref{eq:2moderdm} and \eqref{eq:bos_exp_mode}. When assessing entanglement and correlations of mode-reduced density operators, it becomes necessary to take into account that some parts of the density operators are not accessible by physical observations. More specifically, super-selection rules prohibit certain (projective) measurements, which in turn constrains the algebra of observables \eqref{eqn:obs}. This shall be discussed in Sec.~\ref{sec:ssr}.
    
	\subsection{$N$-Harmonium}
	The Hamiltonian of the $N$-Harmonium model system in $n$ spatial dimensions reads:
	\begin{equation}
	\label{eqn:ham}
		H =\sum_{i=1}^{N} (\frac{{{\vec{p}}_i}^{\, 2}}{2m}+ \frac{m}{2}\vec{x}_i^T \Omega \vec{x}_i) \\ + \frac{K}{2} \sum_{1\leq i< j\leq N}(\vec{x}_i-\vec{x}_j)^2,
	\end{equation}
	where $\Omega = diag(\omega^{1}, \cdots, \omega^{n})$ parametrises the harmonic trap strength and $K$ parametrises the particle-particle interactions; $(\vec{x}_1,\ldots,\vec{x}_N)$ and $(\vec{p}_1,\ldots,\vec{p}_N)$ denote particle positions and their respective momenta. Coupling can be attractive $K < 0$  or repulsive $K > 0$. One requires $\frac{-\Omega^2}{m N}<K$  for the existence of bound states. The Hamiltonian can be separated and diagonalised by a suitable coordinate transformation (cf.~Ref.~\cite{Schilling2013} for further details).

	The bosonic ground state has the profile of a Gaussian and reads: 
	\begin{equation}
	\label{eq:bos-gs}
	   \Psi ^{(b)}(\vec{x}_1, \dots, \vec{x}_N)= \mathcal{N} e^{\vec{X}^T\boldsymbol{B}\vec{X}-\sum_{i=1}^N(\vec{x}^{T}_i\boldsymbol{A}\vec{x}_i)}
	\end{equation} with $\vec{X}\equiv \frac{1}{N}(\vec{x}_1+\ldots \vec{x}_N)$ representing the centre-of-mass vector, and constants $\boldsymbol{A}\equiv \mbox{diag}(A^{(1)},\ldots,A^{(n)})$, where $A^{\alpha} \equiv \frac{1}{2(\tilde{l}^{(\alpha)})^2}$ and $\boldsymbol{B}\equiv \mbox{diag}(B^{(1)},\ldots,B^{(n)})$, where $B^{(\alpha)}\equiv  \frac{N}{2}\left(\frac{1}{(\tilde{l}^{(\alpha)})^2}
	-\frac{1}{(l^{(\alpha)})^2}\right)$, $\tilde{\omega}^{(\alpha)}\equiv \sqrt{(\omega^{(\alpha)})^2+\frac{NK}{m}}$,
	$\tilde{l}^{(\alpha)} \equiv \sqrt{\frac{\hbar}{m \tilde{\omega}^{(\alpha)}}}$, $ l^{(\alpha)} \equiv\sqrt{\frac{\hbar}{m\omega^{(\alpha)}}}$, and $\mathcal{N}$ represents a normalisation constant.
	
    The fermionic ground state is found to equate to a product of a Slater determinant of Hermite functions and an exponential factor of the centre-of-mass coordinate. This represents a product of a determinant of polynomials (the Vandermonde determinant, which is anti-symmetric under the exchange of particles) and the bosonic ground state (cf.~Ref.~\cite{Tennie2016}). More explicitly, the $N$-fermion ground state $\Psi^{(f)}$ of the Harmonium model is given by:
	\begin{equation}\label{eq:fermgs}
	\Psi^{(f)}(\vec{x}_1, \dots , \vec{x}_N)=\mathcal{N}\,\cdot \left| \begin{array}{c c c}
	\phi_{\bm{\mu}_1}^{(\mathbf{\tilde{\mathbf{l}}})}(\vec{x}_1) & \cdots & \phi_{\bm{\mu}_1}^{(\mathbf{\tilde{\mathbf{l}}})}(\vec{x}_N) \\
	\vdots& \, &\vdots\\
	\phi_{\bm{\mu}_N}^{(\mathbf{\tilde{\mathbf{l}}})}(\vec{x}_1)&\cdots&\phi_{\bm{\mu}_N}^{(\mathbf{\tilde{\mathbf{l}}})}(\vec{x}_N) 
	\end{array}
	\right| \cdot e^{\vec{X}^T\boldsymbol{B}\vec{X}}\, .
	\end{equation}
    Here, the $\phi_{\bm{\mu}_k}^{(\mathbf{\tilde{\mathbf{l}}})}$ represent products of Hermite functions of the relevant spatial dimensions: $\phi_{\bm{\mu}}^{(\mathbf{\tilde{\mathbf{l}})}}(\vec{x}_i)=\prod_{\alpha=1}^{n} \phi_{\mu^{\alpha}}^{(\tilde{l}^{\alpha})}(x^{\alpha}_{i})$, where $\phi_{\mu^{\alpha}}^{(\tilde{l}^{\alpha})}(x^{\alpha}_{i})$ are Hermite functions of degree $\mu^{\alpha}$ and natural length scale $\tilde{l}^{\alpha}$. 
  
	The quantum number vectors $\bm{\mu_1}, \cdots, \bm{\mu_N}$ are determined by a form of an exclusion principle while minimising the total energy. In this instance, one should  consider the spin degrees of freedom to be effectively frozen out by imposing a strong external magnetic field which aligns them. Consequently, in spin-aligned Harmonium no pair of $\bm{\mu_i}$ can be identical.
	Note that $l^{(\alpha)}$ is an intrinsic length scale and one can set $l^{(1)}= 1$ without loss of generality. The lengths $\tilde{l}^{(\alpha)}$ are then determined by the dimensionless coupling constants:
	\begin{equation}\label{eq:kappa}
	\kappa^{(\alpha)} = \left( \frac{l^{(\alpha)}}{\tilde{l}^{(\alpha)}}\right)^4-1.
	\end{equation}
	$\kappa^{(\alpha)}$ is bounded, as one requires $\kappa^{(\alpha)} > -1$ for bound state solutions. We can thus express the bosonic and fermionic ground states as a function of the dimensionless constant $\kappa^{(\alpha})$.

  		\subsection{Super-selection Rules}
  		\label{sec:ssr}
  	The study of entanglement in non-relativistic systems of indistiguishable particles has motivated much discussion with regards to determining the part of the physical state space, which is in principle experimentally accessible. 

   It is widely accepted that certain fermionic and bosonic states are un-physical. More specifically, it appears impossible to form coherent superpositions of states possessing different values of a conserved quantity such as the (global) particle number. For instance, a coherent superposition of a one-electron and two-electron state $\ket{\psi}= \alpha\ket{e^-} + \beta\ket{e^-, e^-}$ cannot be realised in laboratories. This is an example of states that are prohibited by the global particle number super-selection rule (g)N-SSR: for certain types of particles (e.g.~massive fermions), isolated systems have a fixed and conserved total number of particles. Mathematically, this amounts to the particle number operator commuting with the density operator of the system, $[\hat{N}, \rho] = 0$. For systems which are partitioned into subsets $A$ and $B$ with respect to a particular mode, this further implies that states must be of the form \cite{dalton2016quantum}:
  		\begin{equation}
  		    \ket{\Phi_N} = \sum_k \alpha_{k,N} \ket{k}_A \otimes \ket{N-k}_B,
  		\end{equation}
  		where $k \in [0,1]$ for fermions, whereas $k \in [0,N]$ for bosons. 
    In this paper, we consider ground states of bosonic and fermionic Harmonium which have a fixed particle number and are subject to the (g)N-SSR. In addition to the (g)N-SSR, local parity and particle numbers with regards to a particular mode sub-system, further constrain the physically realisable density operators. This will be discussed subsequently for both fermions and bosons.

  	\subsubsection{Fermions}
  		    \label{sec:fer_ssr}
  
 Within the realm of experimental realisation, two constraints relevant for the construction of mode reduced density operators are known for fermionic systems: the local parity-super-selection rule (P-SSR) \cite{wick_52} and the local particle number-super-selection rule (N-SSR) \cite{wick_70}. 
 
 P-SSR precludes a superposition of even and odd-number fermion states. Under local conservation of parity $q$ (P-SSR), observables $\hat{O}_{A/B}$ must satisfy:
\begin{equation}
  \label{eqn:pssr}
  	\hat{O}_{A/B}^P = \sum_q P_q  \hat{O}_{A/B} P_q,
  \end{equation}
where the $P_q$ represent projectors onto subspaces of definite parity number. Therefore, the observables $\hat{O}_{A/B}^P$ are block-diagonal and the algebra of observables decomposes into an even an odd parity sector $\ACAL_{ee} + \ACAL_{oo}$. Note that a violation of local parity super-selection would have wide-ranging consequences such as super-luminal signalling \cite{Vidal, Ding2020, Debarba2020} and is therefore considered impossible.

   Local N-SSR is an even more stringent constraint and it requires the local observables $\hat{O}_{A/B}^N$ to be block-diagonal in the particular fermion-N-particle number sectors:
  \begin{equation}
  \label{eqn:nssr}
  	\hat{O}_{A/B}^N = \sum_{N \geq0} P_N \hat{O}_{A/B} P_N,
  \end{equation}
  where the $P_N$ denote the projectors onto the local N-fermion subspace. 

  Local number super-selection has been the subject of many discussions. Though it cannot be related to fundamental laws of nature such as no-super-luminal-signalling, it applies to most practical scenarios. Recently, works have suggested that the status of N-SSR is challenged by the existence of pure states with indefinite particle numbers (e.g. Majorana fermions, cf.~Ref.~\cite{maj_mourik}). One can however say, that the creation of useful coherent superpositions of states with different numbers of fermions is incredibly challenging. Within the realm of current experimentation, N-SSR tends to be satisfied \cite{Debarba2020} as any results claiming to realise N-SSR violating states have not been reproduced \cite{Frolov2021}.
  
    \subsubsection{Bosons} \label{sec:bos_ssr}
    
    Unlike fermions, bosonic systems do not exhibit a local parity super-selection rule; however, a bosonic local N-SSR for the construction of physical density operators has been scrutinuously discussed (cf.~\cite{dalton2016quantum, bartlett}). In general, one must discern systems of massive and massless bosons. For reasons similar to the ones outlined in Sec.~\ref{sec:fer_ssr}, a local N-SSR should be applied to massive bosons; in particular, the realisation of coherent superpositions of massive bosons with different local particle numbers has proven to be extremely challenging. Thus, the local observables, allowed under local N-SSR, are those which are block-diagonal with respect to the bosonic N-particle number sectors. This algebra is constructed analogously to its fermionic counterpart defined in Eq.~\eqref{eqn:nssr}.
    
    The introduction of a local N-SSR is motivated by the fact that there is no known methodology to easily prepare massive bosonic states which violate local N-SSR \cite{dalton2016quantum}. Note that the status of local N-SSR, particularly for massless bosons, is still subject to much debate \cite{bartlett, photon-ssr, ref_frame-bartlett}, hinging upon a choice of phase reference. However, the status of N-SSR for massless bosons is not of further relevance to the Harmonium model which describes massive particles.

    	\subsection{Symmetries and Entanglement}
\label{sec:symmetries}
     In this section, we briefly state a theorem which becomes instrumental when calculating entanglement. Genererally, due to the possibly high dimensionality of the state space, it is very challenging to explicitly carry out the minimisation in Eq.~\eqref{eqn:relent}. However, for states which exhibit local unitary symmetries, the computation simplifies significantly \cite{werner_symmetries}: 
     \begin{theorem}
        If a density operator $\rho$ is invariant under local unitary operations U(g), representing elements of a group $g \in G$, with generator $\mathcal{T}_{G}$, s.t. ${U^{\dagger}(g) \rho U(g) = \rho}$, then the closest separable state with respect to $\rho$, $\sigma^{*}$, possesses the same symmetries as $\rho$, such that $\mathcal{T}_{G}(\sigma*) = \sigma*$.
    \end{theorem}
    Without proving the statement which is done elsewhere (cf.~\cite{Ding_thesis}), we merely reiterate that the local unitaries are not entanglement generating. In this paper, we focus on the symmetries of local parity and particle number, as defined in Eqs.~\eqref{eqn:pssr} and \eqref{eqn:nssr}. In some cases, these symmetries even permit an analytical evaluation of the relative entropy of entanglement.

	\section{Results}
	\label{sec:results}

		
We demonstrate the impact of particle number and parity constraints on entanglement in fermionic and bosonic Harmonium. The ground state of Harmonium is a pure and fixed particle number state; it therefore obeys the global N-SSR. Due to local super-selection rules, not all parts of mode-reduced density operators are accessible by physical measurements (cf.~Sec.~\ref{sec:ssr}). As a consequence, care must be taken when determining the amount of physically meaningful entanglement. Our results show a striking impact of super-selection rules and reveal the difference between `apparent' and physically accessible entanglement in Harmonium. Since the Harmonium model of harmonically interacting and trapped particles is considered as a reasonable first approximation to systems such as quantum dots (cf.~Refs.~\cite{johnson91, teichmann2013}), our results demonstrate the need to re-evaluate the notion of entanglement in various other continuous systems.

Hermite functions constitute a natural single-mode basis of the Harmonium model. In what follows, we shall use bases constructed out of Hermite functions $\phi_{m}^{(\tilde{l})}$ of natural length scale $\tilde{l}$ (cf.~\ref{section:hermite}). Note that by Eq.~\eqref{eq:kappa}, selecting a different arbitrary length scale $\bar{l}$ would merely amount to `stretching' or `quenching' the $\kappa$ axes in the correlation and entanglement measure plots below. Using the basis of Hermite functions, Slater determinants (and permanents) will be abbreviated by their respective mode number, e.g.~
\begin{equation} \label{eqn:slnotation}
	 f^{\dagger}_0 f^{\dagger}_1 f^{\dagger}_2 \ket{\Omega}\equiv\ket{\phi_{0}, \phi_{1}, \phi_{2}} \equiv \ket{m_0,m_1,m_2}
\end{equation}
for ease of notation.

	\subsection{Fermions}
	\label{sec:ferm_mode_ent}
 	In the presence of an external magnetic field, an additional Zeeman term $\HC_{B} = \frac{c_{S}}{\hbar} \sum_{i}^{N}\hat{S_i} \cdot \vec{B_{ext}}$ must be added to the Harmonium Hamiltonian \cite{Tennie2016}, where $\hat{S_i}$ represents the spin angular momentum operator of the $i$-th particle, $c_S$ is a constant and $B_{ext}$ represents the external magnetic field. The additional Zeeman term commutes with the $N$-Harmonium Hamiltonian and leads to a splitting of the single-particle states to $\bar{\phi}_{i}(x;\uparrow \! \! / \! \! \downarrow) = \phi_{i}(x) \otimes \ket{\uparrow \! \! / \! \!  \downarrow}$. In spin-ful Harmonium, the ground state can be determined analogously to Eq.~(\eqref{eq:fermgs}) by replacing $\phi$ with $\bar{\phi}$. Note that the relative strength and orientation of the magnetic field determines the ground-state configuration. 
  
  In the regime of strong magnetic fields, i.e. ${\abs{B_{ext}} \gg \frac{\hbar}{c_s}\omega\sqrt{1+\kappa}}$, the system's spin degree of freedom will be quenched and all spins aligned. The resulting ground state equals the ground state of spin-less Harmonium.

  For smaller field strengths, the individual spins will only partially be aligned and the configuration of the ground state is modified accordingly. For instance, if $\frac{\hbar}{c_s}\omega\sqrt{1+\kappa} < \abs{B_{ext}} < \frac{2\hbar}{c_s}\omega\sqrt{1+\kappa}$, the  ground-state eigenfunction for 3-Harmonium in one spatial dimension ($n=1$) reads: 
  	\begin{equation} \label{eq:spings}
  		\NCAL_{0}^{s} \begin{vmatrix}
      \phi_{0}^{\tilde{l}}(x_1) \otimes \ket{\downarrow} & \phi_{0}^{\tilde{l}}(x_2) \otimes \ket{\downarrow}  & \phi_{0}^{\tilde{l}}(x_3)  \otimes \ket{\downarrow}  \\ 
      \phi_{1}^{\tilde{l}}(x_1) \otimes \ket{\downarrow}       &  \phi_{1}^{\tilde{l}}(x_2) \otimes \ket{\downarrow} &  \phi_{1}^{\tilde{l}}(x_3) \otimes \ket{\downarrow}   \\
       \phi_{0}^{\tilde{l}}(x_1) \otimes \ket{\uparrow}     & \phi_{0}^{\tilde{l}}(x_2)  \otimes \ket{\uparrow}   & \phi_{0}^{\tilde{l}}(x_3)  \otimes \ket{\uparrow}  \\ 
    \end{vmatrix} \cdot e^{\vec{X}^T\boldsymbol{B}\vec{X}}.
  	\end{equation}

	\subsubsection{Mode Entanglement}
	
\begin{figure}[h] 
\includegraphics[width=0.52\textwidth]{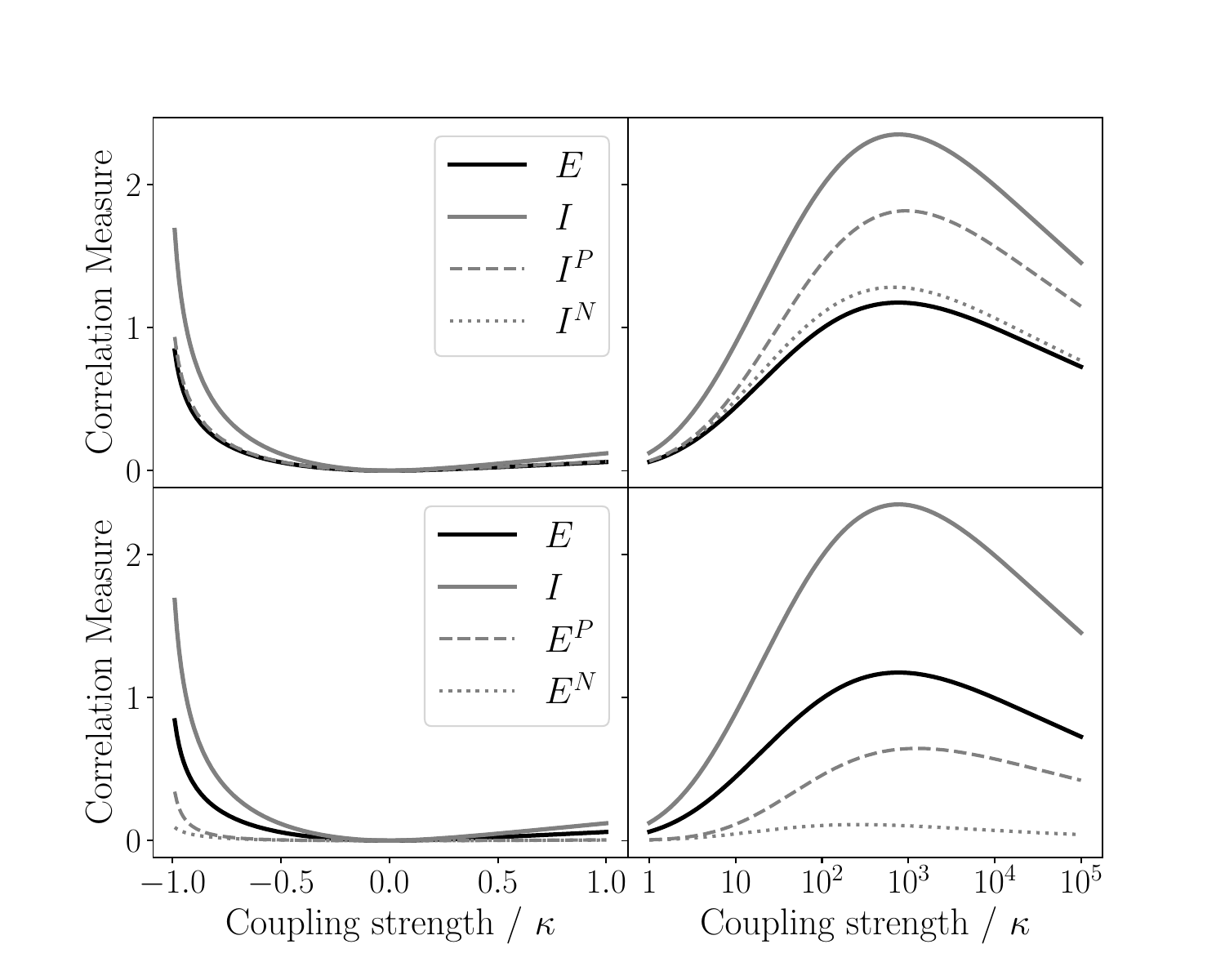}
\includegraphics[width=0.52\textwidth]{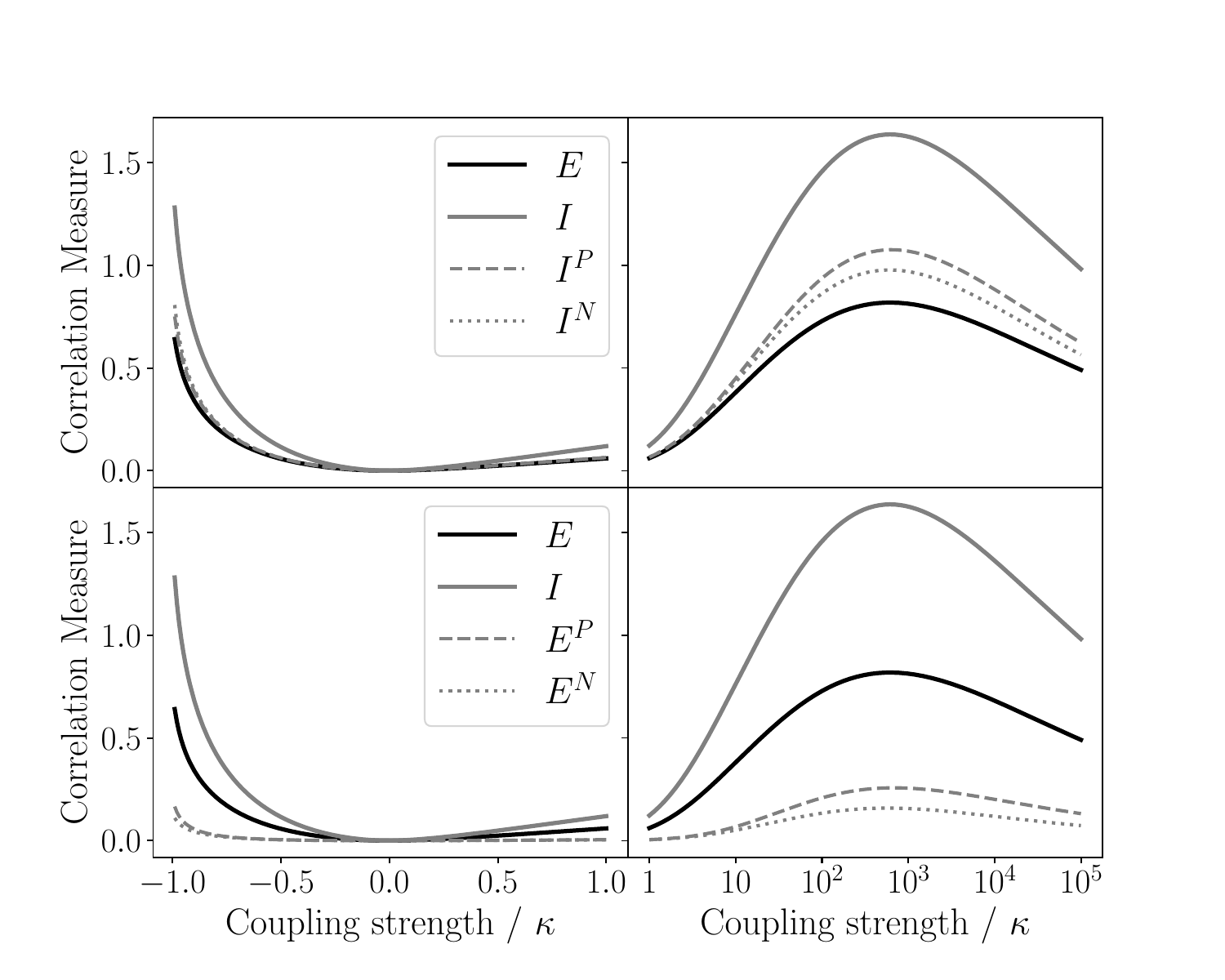}
\caption{Correlation and entanglement of spatial modes $\phi_0$ (top), $\phi_2$ (bottom) in one-dimensional $3$-fermion Harmonium as a function of the relative particle-particle interaction strength $\kappa$. Notably, physically accessible correlation ($I^{N/P}$) and entanglement ($E^{N/P}$) are significantly smaller by comparison to the measures calculated without invoking SSR's ($E$ and $I$).}
\label{fig:spinmode02}
\end{figure}

\begin{figure}[h] 
\includegraphics[width=0.52\textwidth]{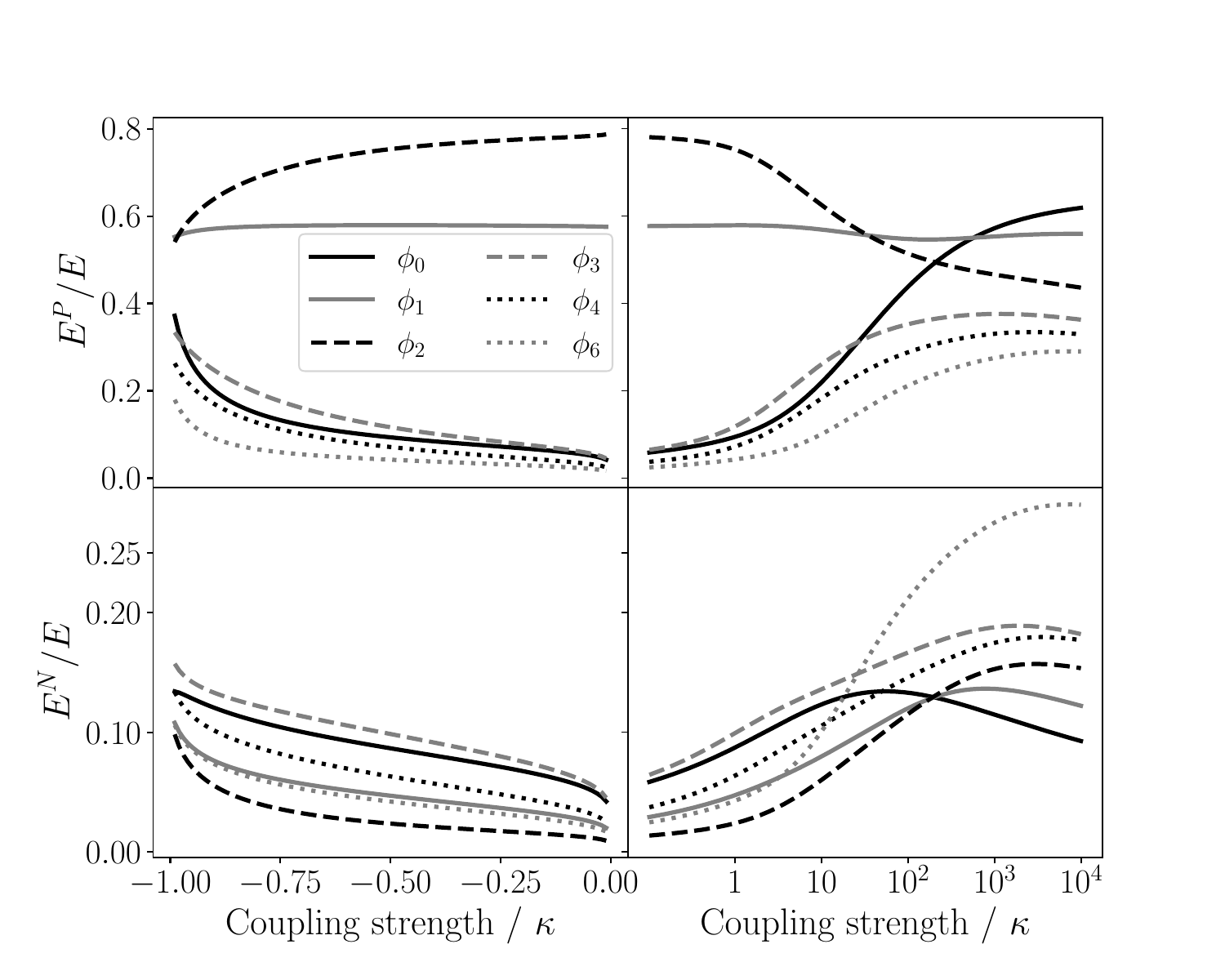}
\caption{Relative amount of physically accessible entanglement $E^{P/N}/E$ of spatial modes $\phi_0$ to $\phi_6$ as a function of the coupling strength $\kappa$ in one-dimensional 4-fermion Harmonium.}
\label{fig:epnssr_frac}
\end{figure}

\begin{figure}[h] 
\includegraphics[width=0.52\textwidth]{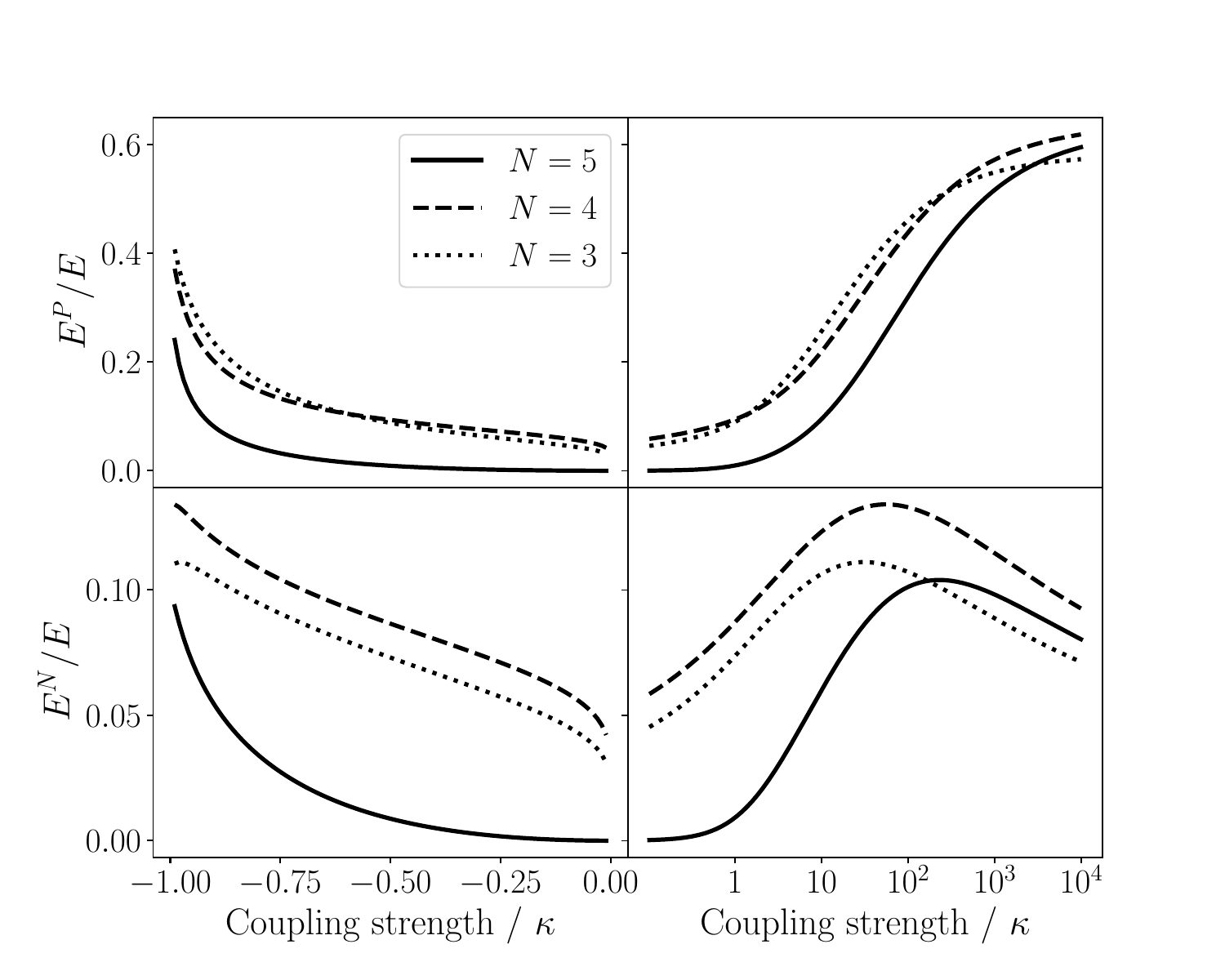}
\includegraphics[width=0.52\textwidth]{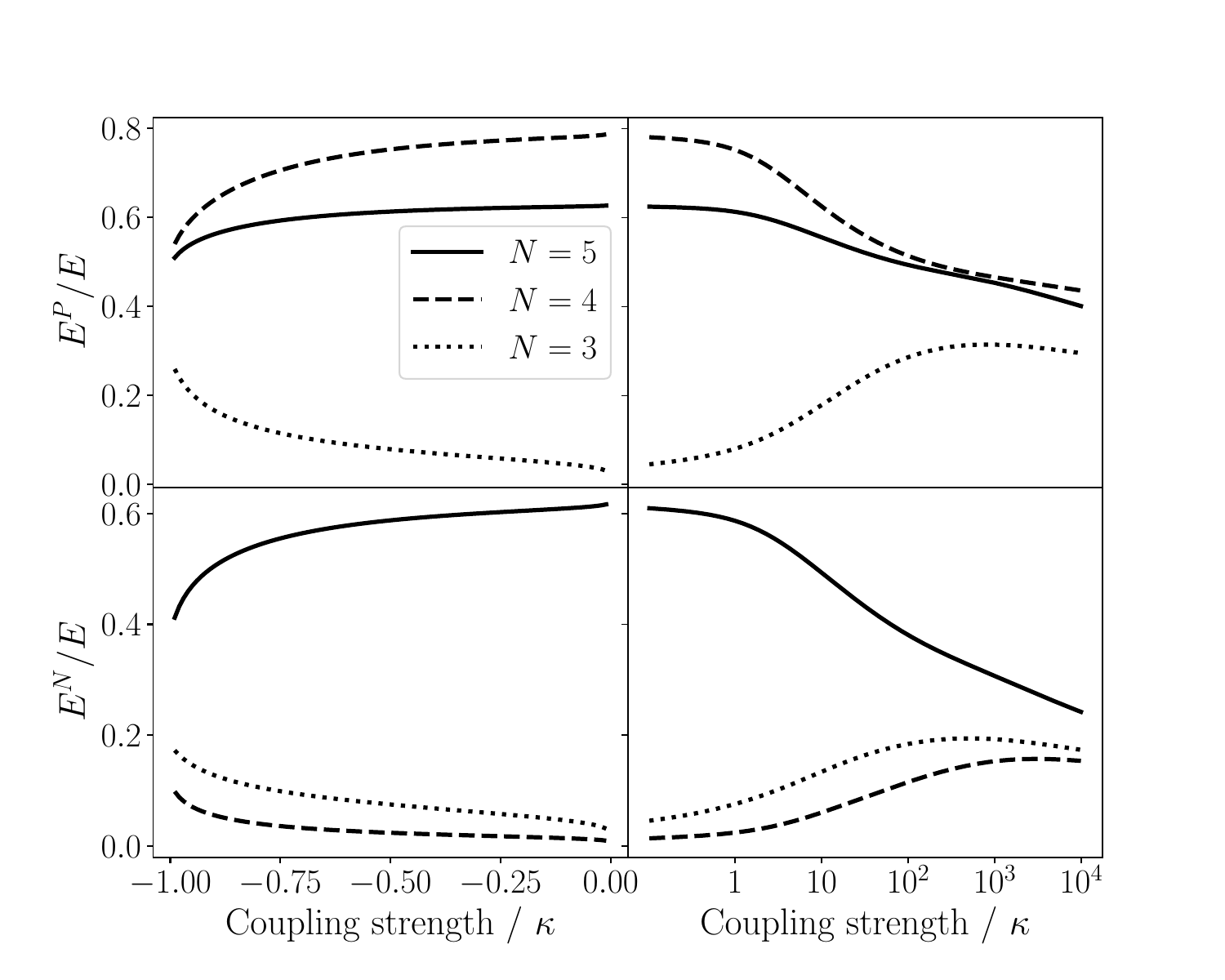}
\caption{Relative amount of physically accessible entanglement $E^{P/N}/E$ of spatial modes $\phi_0$ (top) and $\phi_2$ (bottom) as a function of the coupling strength $\kappa$ for one-dimensional Harmonium of different fermion numbers $N=3,4,5$.}
\label{fig:epnssr_particles}
\end{figure}

	In the presence of a weak external field in spin-ful Harmonium, the mode reduced density operator of spatial mode $m$ becomes diagonal in the local basis $\lbrace \ket{\Omega}, \ket{m_\uparrow}, \ket{m_\downarrow}, \ket{m_\uparrow, m_\downarrow} \rbrace$:
\begin{equation}
\label{eq:onemoderdospin}
	\rho_{ \{ m_\uparrow, m_\downarrow \}} = \text{diag}(q_0, q_1, q_2, q_3)
\end{equation}
 The diagonal representation arises from the spin symmetries and fixed total particle number of the Harmonium ground state (cf.~Eq.~\eqref{eq:2moderdm}). The structure of the Schmidt-decomposed $N$-particle state reads:
\begin{equation}\label{eq:SchmidtDecomposition}
	\begin{aligned} 
	\ket{\psi}=& \sqrt{q_0}\ket{\Omega} \otimes \ket{\psi_{N}} + \sqrt{q_1}\ket{m_\uparrow} \otimes \ket{\psi_{N-1}} \\
	 &+ \sqrt{q_2}\ket{m_\downarrow)}\otimes \ket{\psi_{N-1}'} + \sqrt{q_3}\ket{m_\uparrow, m_\downarrow)}\otimes \ket{\psi_{N-2}}.
	\end{aligned}
\end{equation}

In our analysis, we have determined the amplitudes $\sqrt{q_i}$ analytically. This has become possible since the integral expressions for expectation values of fermionic ladder operators (cf.~\ref{sec:expval_int}) amount to integrals of a Gaussian factor multiplied by a polynomial. The analytic forms of the $q_i$ have then been used to derive an analytic expression for the spatial-mode entanglement and mutual information. Utilising the techniques introduced in Ref.~\cite{Ding2020}, super-selection rules are taken into account by removing the unphysical parts of the reduced density operators. For the reader's reference, we give an account of the explicit expressions for the evaluation of correlation and entanglement measures in ~\ref{sec:calc-f-m-ent}, cf. Eqs.~\eqref{eqn:2modecorrelation}, \eqref{eqn:2modePN-I} and \eqref{eqn:2modePN-E}. Note that the respective measures are labeled as $I,E$ (without super-selection rules), and $I^{N/P},E^{N/P}$ (including the N-SSR, and P-SSR, respectively). Though the analytic form of the $q_i$ contains a large number of terms (cf.~\ref{sec:analyticqi} for the special case of $N=3$ fermions), it allows us to derive an analytic expression for the entanglement measures in an infinite-dimensional interacting many-body system. It should be stressed that this itsself is a remarkable result.

\begin{figure}[h] 
\includegraphics[width=0.5\textwidth]{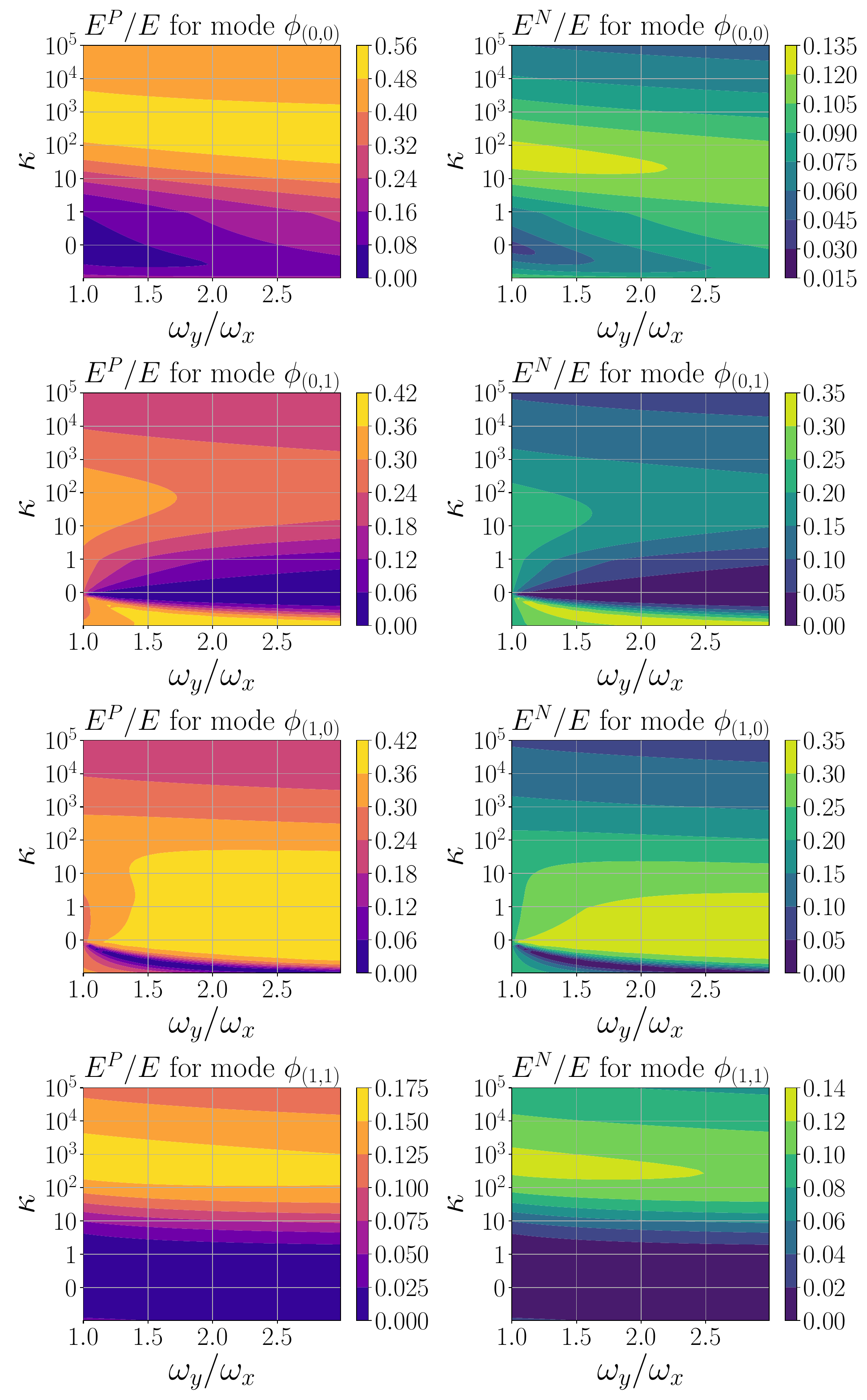}
\caption{Relative amount of physically accessible entanglement $E^{P/N}/E$ in two-dimensional $4$-fermion Harmonium as a function of trapping frequency detuning $\omega_y/\omega_x$ and coupling strength $\kappa$. In two spatial dimensions, the basis of spatial modes consists of products of two Hermite functions, i.e. $\phi_{(i,j)}=\phi_{(i)}\phi_{(j)}$. Note that $\omega_y/\omega_x$ is bounded from above by $3$. For relative frequency detunings exceeding this value, the system has a different ground state which is effectively one-dimensional. Colour online.}
\label{fig:econtour}
\end{figure}

The broad hierarchy and variation of the spatial mode entanglement and correlation measures with $\kappa$ (cf.~Eq.~\eqref{eq:kappa}) is qualitatively similar for various modes as shown in Fig.~\ref{fig:spinmode02}. As an example, this figure displays correlation and entanglement of modes $\phi_0$ and $\phi_2$ in one-dimensional 3-fermion Harmonium. Unsuspectedly, all correlation and entanglement measures assume a local maximum at a finite positive value of $\kappa$ before monotonically decreasing with further growing interaction strengths. This is a peculiar feature of the Harmonium system, insofar as one would expect mode entanglement to grow monotonically with the particle-particle interaction strength $\kappa$, bounded, only by the theoretical maximal entanglement of $E_{max}=ln(4)$. Furthermore, one observes that the amount of physically accessible entanglement decreases significantly once SSR's are taken into account; (in Fig.~\ref{fig:spinmode02} compare $E$ and $I$ with $E^{P/N}$ and $I^{P/N}$). Henceforth, we shall therefore evaluate and display the relative amount of physically accessible entanglement $E^P/E$ and $E^N/E$.\footnote{At $\kappa = 0$ the entanglement $E$ vanishes and the fractions  are not defined. Thus, we shall provide the graphs of $E^P/E$ and $E^N/E$ with analytic continuation at zero interaction strength. }  

Fig.~\ref{fig:epnssr_frac} shows a comparison of the relative amount of physically accessible entanglement between different spatial modes in one-dimensional 4-fermion Harmonium. With the exception of modes $\phi_1$ and $\phi_2$, the relative amount of entanglement accessible under P-SSR approaches zero for small interaction strengths. For large $\kappa \gtrsim 10^3$, $E^P/E$ is hierarchically ordered with the mode number, which is not the case in the intermediate regime $\kappa \approx 1 \ldots 100$. Interestingly, the relative amount of entanglement of mode $\phi_1$ accessible under P-SSR remains approximately constant, while for $\phi_2$ it assumes a maximum around small interaction strengths. A slightly different behaviour is observed for the relative amount of entanglement accessible under N-SSR. For all modes, a local minimum is located around vanishing $\kappa$. Remarkably, the magnitudes of the maxima at large $\kappa$ grow with increasing mode number.\footnote{One might speculate that this result is a reminiscence of the impact of SSRs becoming negligible in the many-copy limit, cf.~Ref.~\cite{fluffy_bunny}.} For instance, $E^N$ of mode $\phi_6$ at around $\kappa\approx 10^4$ accounts for approximately $30\%$ of the entanglement $E$, the highest value of all modes under consideration. It should be noted that the relative amount of entanglement makes no statement on its absolute value, which decreases significantly for higher mode numbers. For the sake of completeness, additional plots of the entanglement measures for different particle numbers are shown in \ref{sec:AppSingleModeFerm}.

The variation of spatial mode entanglement in one-dimensional Harmonium with the particle number is shown in Fig.~\ref{fig:epnssr_particles}. For mode $\phi_0$, the relative amount of entanglement under P-SSR and N-SSR assumes a minimum around vanishing $\kappa$. Throughout, $E^N$ is smaller than $E^P$ and only amounts to less than approximately $10\%$ of the total entanglement. For mode $\phi_2$, the relative amount of entanglement assumes a maximum around vanishing $\kappa$ for some particle numbers while being minimal for others. Note that no clear hierarchy of the relative amount of accessible entanglement with the particle number is observed, in contrast to the particle entanglement results of Ref.~\cite{Benavides-Riveros2014}. 

In two-dimensional Harmonium, the form of the ground state is parametrised by the detuning or quench of the trapping potential frequencies $\omega_y/\omega_x$ and the relative interaction strength $\kappa$. In Fig.~\ref{fig:econtour} the relative amount of physically accessible entanglement is shown for 4-fermion Harmonium. Note that for detunings exceeding a certain threshold value, e.g.~$\omega_y/\omega_x = 3$ for $N=4$ particles, the structure of the two-dimensional Harmonium ground state changes and the system becomes effectively one-dimensional. Generally, we observe that for fixed frequency detuning the amount of accessible entanglement qualitatively behaves similarly to the case of one-dimensional Harmonium and assumes a local maximum at finite positive $\kappa$. For fixed interaction strength, increasing the quench of the trap generally increases the relative amount of accessible entanglement. As in the case of one-dimensional Harmonium, $E^N$ is always smaller than $E^P$.

	\subsubsection{Mode-Mode Entanglement}
In order to evaluate entanglement between two spatial fermionic modes, one has to consider the two-mode reduced density operator. For spin-$1/2$ fermions, this operator has a $16 \times 16$ dimensional matrix representation, since each spatial mode can be occupied by two anti-aligned spins. As in the case of spatial-mode entanglement, we gauge the impact of local super-selection rules by computing the entanglement measures with and without projecting the two-mode reduced density operator onto the parity and number eigenstate basis. The representation of such a two-mode-reduced density operator is well known in quantum chemistry \cite{boguslawski_orbital} and becomes block-diagonal in the number state basis for pure states of fixed (global) particle number such as the Harmonium ground state. Note that this implies that Harmonium cannot exhibit any mode-mode correlation between number blocks with different particle numbers.

All coefficients of the two-mode reduced density operator can be computed analytically. Using fermionic anti-commutation relations \eqref{eq:ferm_creat} the matrix elements can be expressed in terms of particle-reduced density operators (cf.~\ref{sec:calc-f-mm-ent}). Equipped with the analytic form of the two-mode reduced density operator and its marginals, the evaluation of the entanglement measures involves a parametrised optimisation to determine the closest separable state (cf.~Eq.~\eqref{eqn:relent}). Though generally challenging, this task simplifies significantly once symmetries are taken into account (cf.~Sec.~\ref{sec:symmetries}). The optimisation is carried out using the PPT criterion \cite{ppt_separability, num_ent} as a necessary condition for separability. When evaluating $E^{P/N}$, this criterion provides an exact value of the entanglement since all matrix blocks are at most of size $2\times 2$. When evaluating $E$, the criterion only provides a lower bound \cite{ppt_violation}. However, since we are interested in gauging the relative amount of physically accessible entanglement this lower bound is sufficient. Furthermore, numerical computations reveal that this lower bound is almost exact. 

\begin{figure}[h] 
\includegraphics[width=0.5\textwidth]{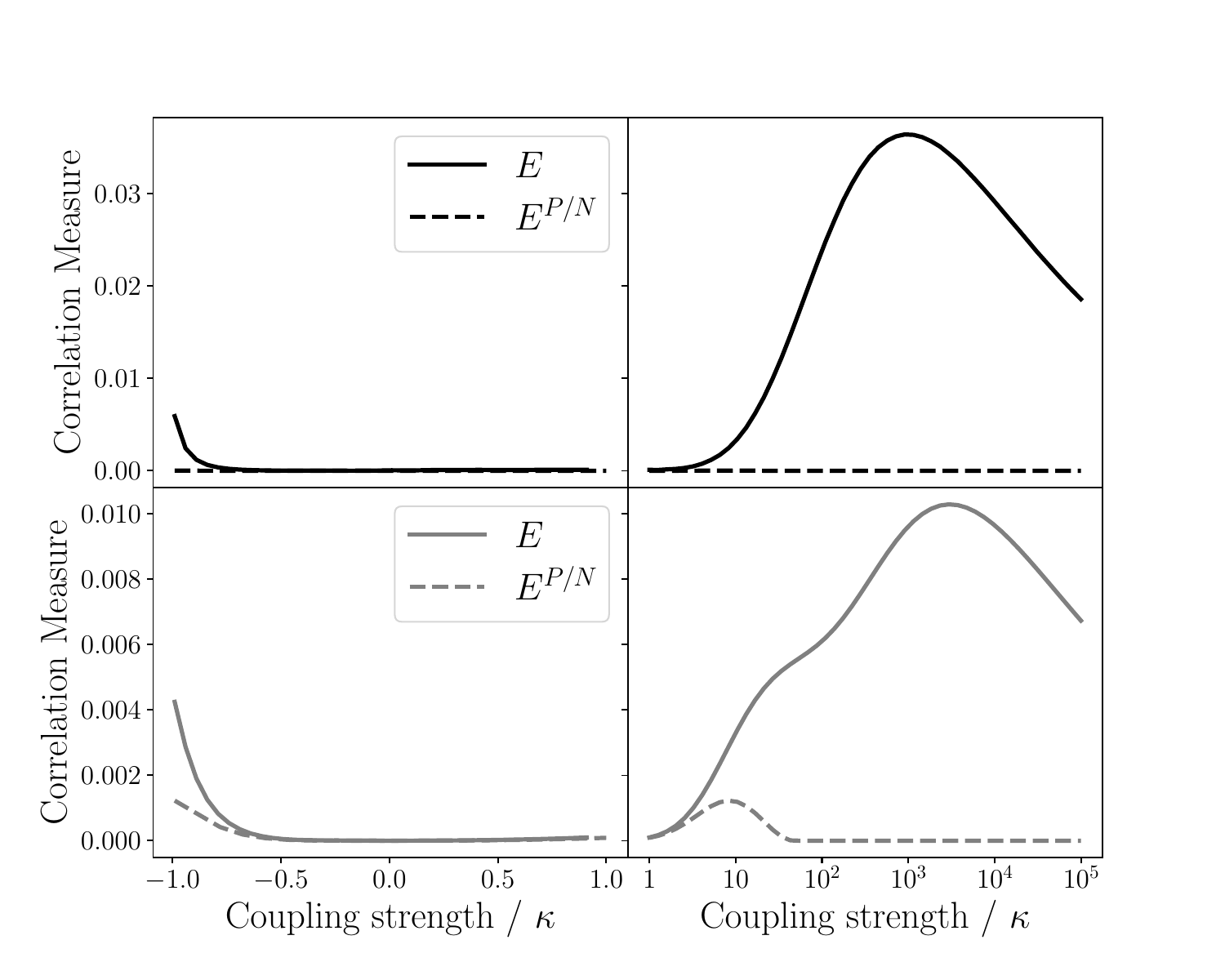}
\caption{Mode-mode entanglement between mode $\phi_0$ and mode $\phi_1$ (top, black), and between mode $\phi_0$ and mode $\phi_2$ (bottom, grey) in 4-fermion Harmonium as a function of $\kappa$. Note that the evaluation of the entanglement measures involves a parametrised minimisation. Here, we have used the PPT criterion \cite{ppt_separability} for separability which yields an exact value for $E^{N/P}$ and provides a lower bound for $E$.}
\label{fig:fermion_modemode01}
\end{figure}

Mode-mode entanglement in $N$-fermion Harmonium is found to behave qualitatively like in the specific case of $N=4$ fermions in one spatial dimension, Fig.~\ref{fig:fermion_modemode01}. We observe that the relative amount of physically accessible entanglement $E^{P/N}$ is smaller  by several orders of magnitude than $E$, cf.~Eq.~ \eqref{eq:modemodeEN} for an explicit definition of $E^{P/N}$. It has a minimum around vanishing coupling strength $\kappa$, followed by an increase in the entanglement to a maximum around $\kappa \approx 10^3$ and a subsequent decrease for larger interaction strengths. Interestingly, mode-mode entanglement between the two modes $\phi_0$ and $\phi_1$ that form the ground state at $\kappa=0$ never exceeds $E\approx 10^{-5}$, even in the regime of strong couplings.

So far, we have assessed the amount of physically accessible entanglement in systems of spin-ful fermions. The case of spin-less fermions can be considered by applying an external magnetic field that aligns all spins. However, due to the resulting symmetries no physically accessible mode entanglement is observed while mode-mode entanglement is similar to the case of mode entanglement for spin-ful fermions.

	\subsection{Bosons}
We now consider the entanglement of modes in bosonic Harmonium with integer spin $S=1$. As for fermions, the Zeeman term defined in Sec. \ref{sec:ferm_mode_ent} is added to the Hamiltonian which causes a splitting of the eigenstates with respect to the magnetic quantum numbers $\sigma_b = \lbrace -1,0,1 \rbrace$ \eqref{eq:spings}. The relative orientation of the magnetic field determines the ground-state spin configuration of bosonic Harmonium. Thus, in the presence of an external magnetic field, however weak it may be, it is energetically favourable to occupy a particular spin mode. This scenario is physically realistic, because all physical systems are subject to some external magnetic field, which will perturb the energy levels of the system. For ease of notation, the spin degree of freedom is omitted in what follows since all spins are assumed to be aligned. This is equivalent to considering massive spin-less bosons.

	\subsubsection{Mode Entanglement}
	
	\begin{figure}[h] 
\includegraphics[width=0.52\textwidth]{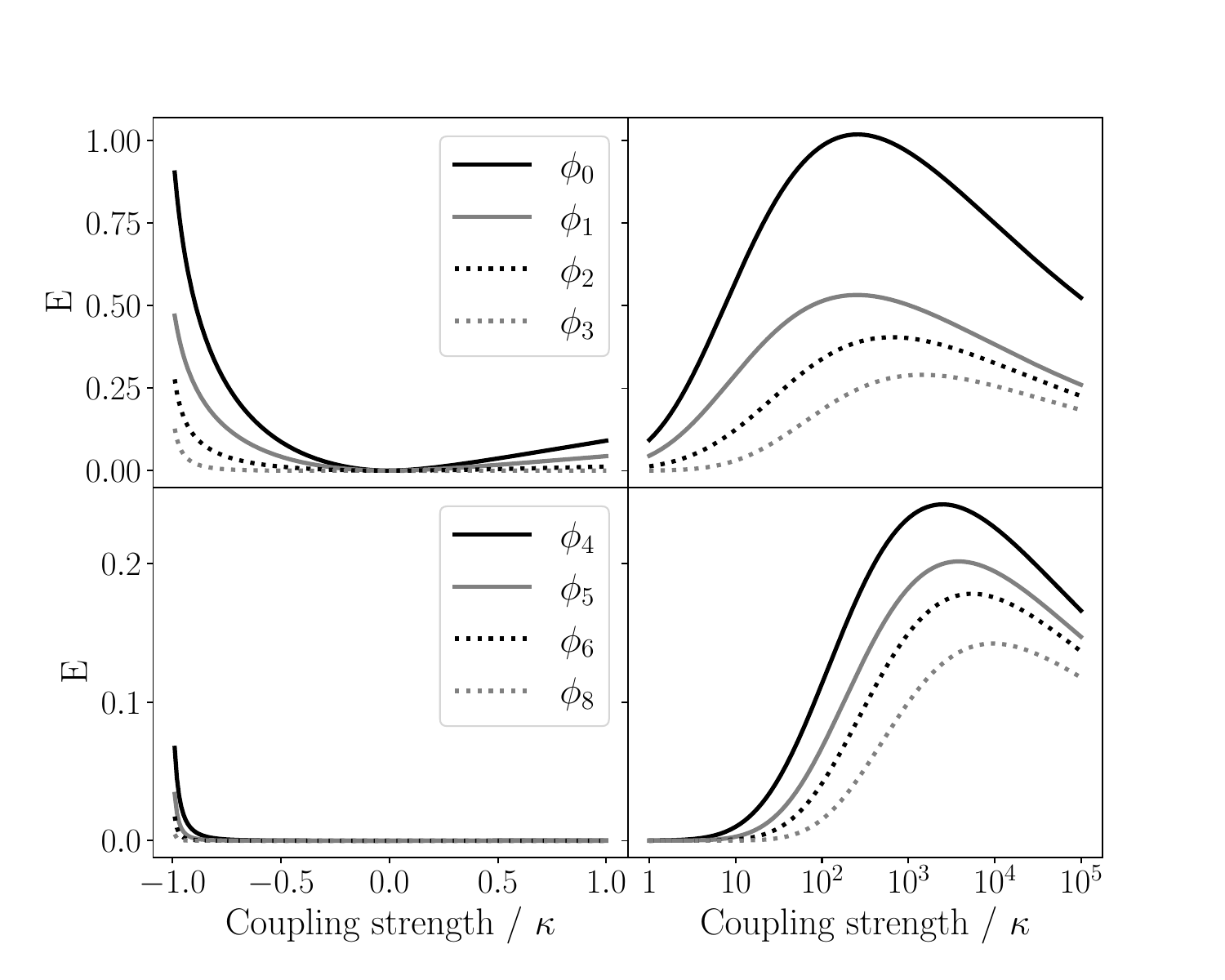}
\caption{E of various spatial modes as a function of $\kappa$ in one-dimensional 3-boson Harmonium.}
\label{fig:bos_me_n3}
\end{figure}

	\begin{figure}[h] 
\includegraphics[width=0.52\textwidth]{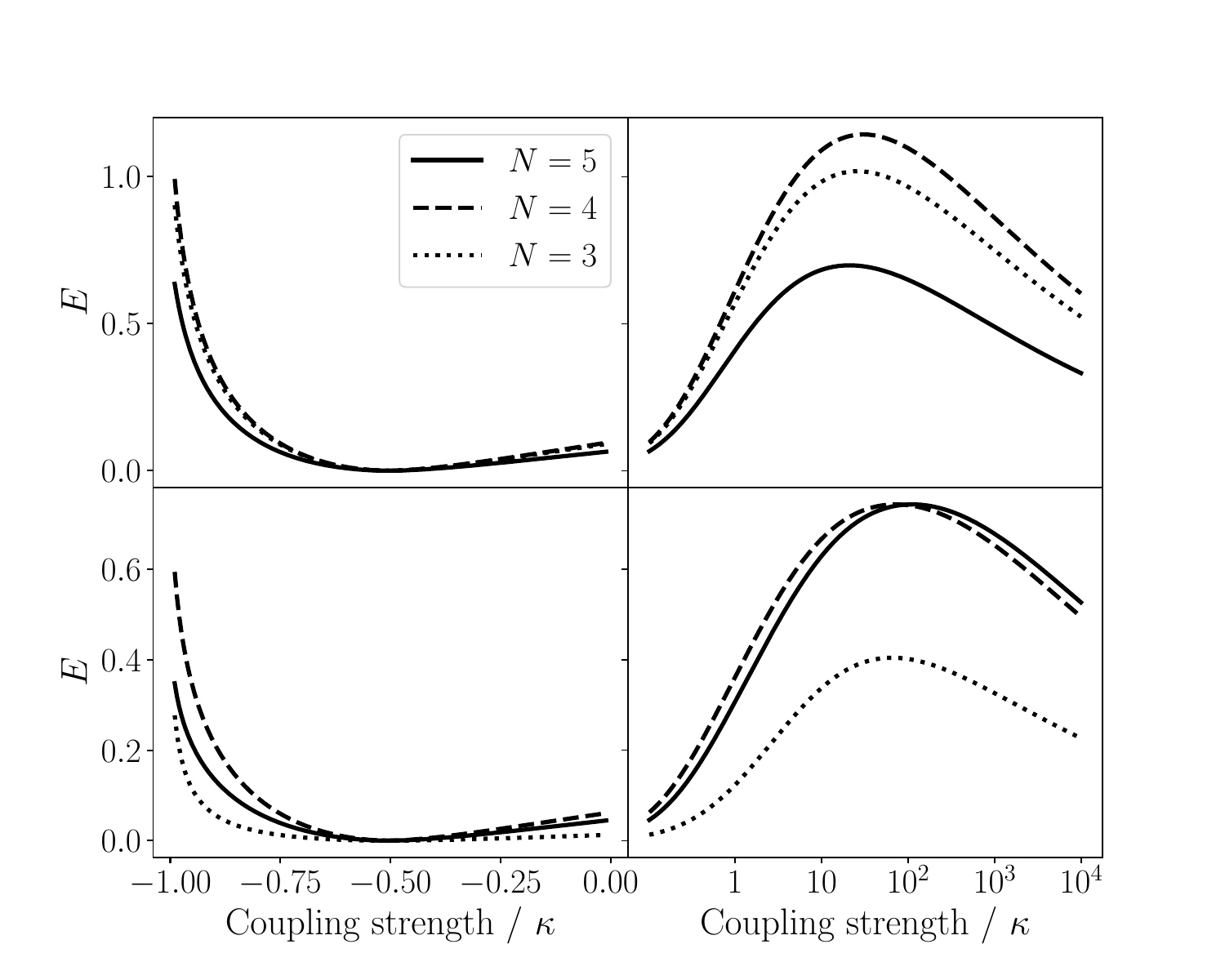}
\caption{E of spatial modes $m_0$ (top) and $m_2$ (bottom) as a function of $\kappa$ for one-dimensional Harmonium with different boson numbers $N=3,4,5$.}
\label{fig:bos_me_nvar}
\end{figure}
    Mode correlation and entanglement are encapsulated in the $(N+1)$ dimensional mode reduced density operator (cf. Sec. \ref{sec:bos-modered}). In the basis $\{(b^\dagger)^i\ket{\Omega}\}^{N}_{i=0}$ this operator becomes diagonal for the pure, fixed particle number ground state of Harmonium.

    For example, for three bosons the mode-reduced density operator of mode $m$ can be parametrised as $\rho_{\{m\}}=diag(p_0,p_1,p_2,p_3)$. Consequently, the Schmidt-decomposition of the three particle state can be written as:

\begin{equation}
\label{eqn:doublemode}
\begin{aligned} 
	\ket{\psi_b^{N=3}}=& \sqrt{p_0}\ket{\Omega} \otimes \ket{\psi_{3}'} + \sqrt{p_1}\ket{m} \otimes \ket{\psi_{2}'} \\
	&+ \sqrt{p_2}\ket{m,m}\otimes \ket{\psi_{1}'} + \sqrt{p_3}\ket{m,m,m}\otimes \ket{\Omega'}.
	\end{aligned}
\end{equation}
Similar to fermions, the matrix elements $p_i$ can be computed by evaluating particle-reduced density operators. The related integrals contain a polynomial multiplied by a Gaussian factor and therefore can be calculated analytically (cf. \ref{sec:expval_int}). 

Due to the Schmidt-decomposition, the mutual information and entanglement read:
\begin{equation} \label{eq:1bosmode}
	E(\rho_b) = \frac{1}{2}I(\rho_b) = - \sum_{k} p_k \ln(p_k).
\end{equation}
Once local N-SSR constraints are imposed, the Schmidt-decomposition of the $N$-boson state also implies that the $N+1$ sectors with fixed local particle number factorise individually. Therefore, we find that the amount of entanglement accessible under N-SSR vanishes, $E^N=0$.

Generally, bosonic mode entanglement behaves qualitatively similar for different particle numbers and modes. In Fig.~\ref{fig:bos_me_n3}, the entanglement of modes $\phi_0$ to $\phi_8$ is displayed as a function of $\kappa$. While vanishing for zero interaction strength, it attains a maximum at finite $\kappa$ in the regime of strong couplings. We also observe a hierarchy of the entanglement which strictly decreases with increasing mode number.

In Fig.~\ref{fig:bos_me_nvar}, the influence of the boson number on mode entanglement is demonstrated for modes $\phi_0$ and $\phi_2$ and $N=3,4,5$. No clear hierarchy with the particle number is observed.

\subsubsection{Mode-Mode Entanglement}

\begin{figure}[h] 
\includegraphics[width=0.5\textwidth]{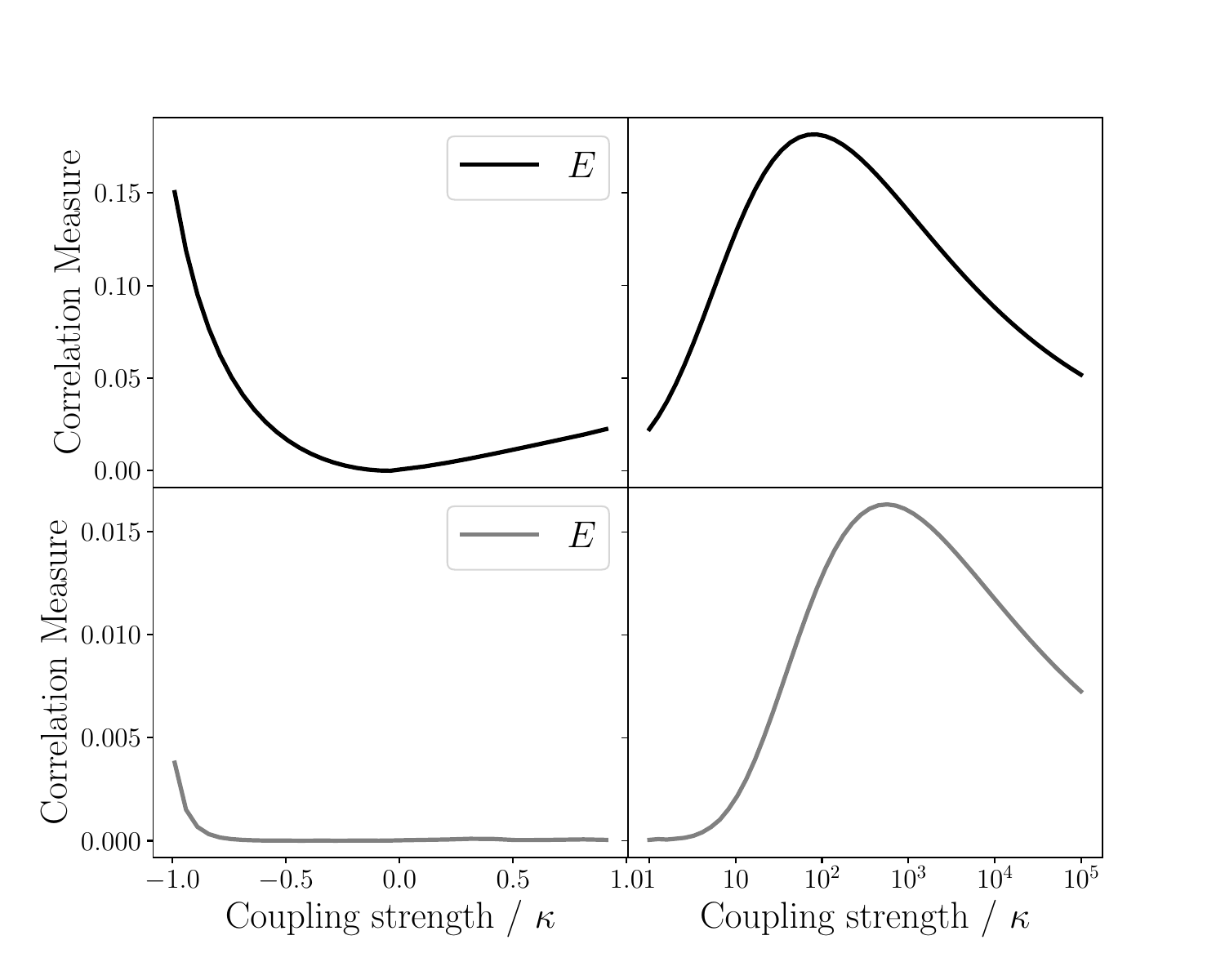}
\caption{Mode-mode entanglement between modes $\phi_0$ and $\phi_1$ (black, top), and between modes $\phi_1$ and $\phi_2$ (grey, bottom) in 3-boson Harmonium as a function of interaction strength $\kappa$.}
\label{fig:bos_mm}
\end{figure}

    To assess bosonic mode-mode entanglement, one has to determine the two-mode reduced density operator. For pure fixed-particle number states, many of its matrix elements in the particle number basis vanish. The example of two bosons is outlined in Tab.~\ref{tab:bos2moderdo}. As above, one can compute the matrix elements of $\rho_{\{m,n\}}$ by evaluating particle-reduced density operators in the basis $\{(b_m^\dagger)^i(b_n^\dagger)^j\ket{\Omega}\}^{2}_{i=0,j=0}$. 
    
    Note that once a local N-SSR is introduced, all local number sectors of $\rho_{\{m,n\}}$ become separable in the fixed local-particle number basis. In Tab.~\ref{tab:bos2moderdo} this amounts to deleting the greyly marked matrix elements. Therefore the entanglement accessible under local N-SSR vanishes, i.e.~$E^N=0$. Equipped with the analytic form of the two-mode reduced density operator, one can calculate the entanglement between two modes by carrying out a constrained minimisation in accordance with Eq.~\eqref{eqn:relent}. Here, we again use the PPT criterion which provides a lower bound for the entanglement \cite{ppt_separability, num_ent}.

\vspace{0.5cm}
\begin{table}[h]
\begin{tabular}{p{1cm}|p{1cm}|p{1cm}p{1cm}|p{1cm}p{1cm}p{1cm}}
\hline
N &
  \multicolumn{1}{c|}{0} &
  \multicolumn{2}{c|}{1} &
  \multicolumn{3}{c}{2} \\ \hline
Modes &
  $\Omega$ &
  \multicolumn{1}{l|}{$m$} &
  $n$ &
  \multicolumn{1}{l|}{$m, m$} &
  \multicolumn{1}{l|}{$m, n$} &
  $n, n$ \\ \hline
$\Omega$ &
  \cellcolor{black} &
   &
   &
   &
   &
   \\ \hline
$m$ &
   &
  \cellcolor[HTML]{000000} &
  \cellcolor[HTML]{9B9B9B} &
   &
   &
   \\ \cline{1-1}
$n$ &
   &
  \cellcolor[HTML]{9B9B9B} &
  \cellcolor[HTML]{000000} &
   &
   &
   \\ \hline
$m, m$ &
   &
   &
   &
  \cellcolor[HTML]{000000} &
  \cellcolor[HTML]{9B9B9B} &
  \cellcolor[HTML]{9B9B9B} \\ \cline{1-1}
$m, n$ &
   &
   &
   &
  \cellcolor[HTML]{9B9B9B} &
  \cellcolor[HTML]{000000} &
  \cellcolor[HTML]{9B9B9B} \\ \cline{1-1}
$n, n$ &
   &
   &
   &
  \cellcolor[HTML]{9B9B9B} &
  \cellcolor[HTML]{9B9B9B} &
  \cellcolor[HTML]{000000} \\ \hline
\end{tabular}
\caption{Elements of the boson 2-mode reduced density operator $\rho_{\lbrace m, n \rbrace}$ with non-zero expectation values, without local N-SSR (grey and black) and with local N-SSR (black).}
\label{tab:bos2moderdo}
\end{table}

Bosonic mode-mode entanglement qualitatively behaves similar for different pairs of modes and particle numbers. Its key features resemble those of fermion mode-mode entanglement. As an example, mode-mode entanglement in one-dimensional $3$-boson Harmonium is shown in Fig.~\ref{fig:bos_mm}. Generally, the amount of entanglement is small and decreases with increasing mode numbers.

\section{Summary and Outlook}

In this paper, we have investigated mode correlation and entanglement contained in the ground state of bosonic and fermionic Harmonium. In particular, we have evaluated the amount of entanglement that remains accessible under physical constraints in form of super-selection rules. Our main findings are that the amount of physically accessible entanglement is considerably smaller than the `fluffy bunny' entanglement which is calculated without taking parity (fermions) and local number (fermions and bosons) constraints into account. Remarkably, it was possible to derive analytic expressions of the entanglement measures which is unusual given that Harmonium is an interacting many-body system on a continuous state space. While considering different modes, different numbers of particles and spatial dimensions, we observed that mode and mode-mode entanglement in Harmonium attain a maximum in the strong coupling regime while approaching zero again in the limit of infinite interaction strength. 

Our investigation of the physically accessible amount of mode-entanglement in Harmonium provides a first analytic insight into mode-entanglement in continuous variable systems. Furthermore, it is of interest to other continuous variable systems whose Hamiltonians have Taylor expansions with leading terms of quadratic order \cite{johnson91}. For such systems, our results provide the basis for a first assessment of the physically meaningful entanglement. This in particular, could be instrumental for gauging the suitability of such systems for quantum information processing tasks, such as quantum teleportation \cite{Olofsson2020} or measurement based quantum computation \cite{ferm_mbqc}. Future research should expand the above concepts to similar Hamiltonians by using state perturbation methods. We remark that our findings indicate a general necessity to re-evaluate the amount of physically meaningful entanglement in systems of bosons and fermions. 

\ack
The authors gratefully thank Lexin Ding, Christian Schilling, Ben Yadin and Axel Kuhn for useful comments and discussions. JOE acknowledges support from the Studienstiftung des Deutschen Volkes. FT acknowledges support from the Royal Society.
The authors declare no conflicts of interest.

\newpage
\bibliography{mode-entanglement.bib}
	
	\appendix
 \section{Data Availability}
 All data and code associated with this publication is available at \cite{github}.
	\section{Hermite Polynomials} \label{section:hermite}
The Hermite polynomials are polynomials that belong to the classical orthogonal polynomials. Different conventions of indexing and prefactors can be found in the literature. For the purposes of this paper the $\nu$-th Hermite polynomial is defined as follows:
\begin{equation}
	H_{\nu}(x)= (-1)^{\nu}e^{x^{2}}\frac{d^{\nu}}{dx^{\nu}}e^{-x^{2}}.
\end{equation}
The first few Hermite polynomials read:
\begin{equation}
	\begin{aligned}
		H_{0}(x)&=1,\\
		H_{1}(x)&=2x,\\
		H_{2}(x) &= 4x^{2}-2, \\
		H_{3}(x)&= 8x^{3}-12x, \\
		H_{4}(x)&= 16x^{4}-48x^{2}+12, \\
		H_{5}(x) &= 32x^{5}-160x^{3}+120x, \\
		H_{6}(x) &= 64x^{6}-480x^{4}+720x^{2}-120.
	\end{aligned}
\end{equation}
The $\nu$-th Hermite function $\phi_{\nu}^{(l)}(x)$ is defined as follows:
\begin{equation}
    \phi_{\nu}^{(l)}(x) = (2^{\nu} \nu !  \sqrt{\pi} l)^{-\frac{1}{2}}e^{\frac{x^2}{2l^2}} H_{\nu}\left(\frac{x}{l} \right).
\end{equation}

	\section{Integrals for expectation values of mode reduced density operators}\label{sec:expval_int}
	The matrix elements of mode- and mode-mode reduced density operators (cf.~Eqs.~\eqref{eq:1moderdm} and \eqref{eq:2moderdm}) can be readily computed using the integral kernel expressions of the one- and two-particles reduced density operators (cf.~definition of $\rho^{(p)}$ in Eq.~\eqref{eq:part_red_dens_operator}). More specifically, the diagonal elements of the mode-reduced density operator for mode $\varphi_m(\vec{x})$ read:
  		\begin{equation}\label{eq:fexp1mode}
  		\begin{aligned}
  		    \left\langle f_m^{\dag}f_m \right\rangle &= \int_{-\infty}^{+\infty} d\vec{x}_1 d\vec{x}'_1 \rho^{(1)}(\vec{x}_1, \vec{x}'_1) \varphi_{m}(\vec{x}_1)\varphi_{m}(\vec{x}'_1), \\
  		    \left\langle f_m f_m^{\dag} \right\rangle &= \sum_{i \neq m} \int_{-\infty}^{+\infty} d\vec{x}_1 d\vec{x}'_1 \rho^{(1)}(\vec{x}_1, \vec{x}'_1) \varphi_{i}(\vec{x}_1)\varphi_{i}(\vec{x}'_1).
  	   	\end{aligned}
  		\end{equation}
  			
  	Similarly, for the fermionic mode-mode reduced density operator \eqref{eq:2moderdm} one obtains an expression for the lower diagonal matrix element:
  		\begin{equation}\label{eq:fexp2mode}
  		\begin{aligned}
  		&\left\langle f_i^{\dag}f_if_i^{\dag}f_i \right\rangle = \\  
        &\int_{-\infty}^{+\infty} d\vec{x}_1 d\vec{x}'_1 d\vec{x}_2 d\vec{x}'_2 \bra{\varphi_i,\varphi_j}\rho^{(2)}(\vec{x}_1, \vec{x}'_1, \vec{x}_2, \vec{x}'_2) \ket{
  		\varphi_i,\varphi_j} \\
        &\text{with } \\
        &\braket{\vec{x}_1,\vec{x_2}|\varphi_i,\varphi_j} = \frac{1}{\sqrt{2}} (\varphi_i(\vec{x}_1),\varphi_j(\vec{x}_2) - \varphi_i(\vec{x}_2),\varphi_j(\vec{x}_1)), \\
  		&\text{and } \\
        &\braket{\varphi_i,\varphi_j|\vec{x}_1',\vec{x}_2'} = \frac{1}{\sqrt{2}} (\varphi_i(\vec{x}_1'),\varphi_j(\vec{x}_2') - \varphi_i(\vec{x}_2'),\varphi_j(\vec{x}_1')).
  		\end{aligned}
  		\end{equation}
    All remaining matrix elements can now be computed using the canonical commutation/anti-commutation relations, e.g.,
        \begin{equation} \label{eq:mrdm_rel}
  		\begin{aligned}
  		    \left\langle f_kf_k^{\dag} \right\rangle = \left\langle f_kf_k^{\dag} f_lf_l^{\dag} \right\rangle +  \left\langle f_kf_k^{\dag}f_l^{\dag}f_l \right\rangle ,\\ 
  		    \left\langle f_k^{\dag}f_k \right\rangle = \left\langle f_k^{\dag}f_k f_lf_l^{\dag} \right\rangle +  \left\langle f_k^{\dag}f_kf_l^{\dag}f_l \right\rangle .
  		\end{aligned}
  		\end{equation}
    Other matrix elements, for instance those given by expectation values of an odd number of ladder operators, evidently vanish for pure states of fixed particle number, such as the Harmonium ground state.
    
    For bosonic Harmonium one can calculate analogous integral expressions by considering the bosonic p-particle reduced density operator and replacing the Slater determinants in Eq.~\eqref{eq:fexp2mode} with permanents of basis functions.

\section{Explicit expressions for fermionic mode entanglement}
\label{sec:calc-f-m-ent}  	
Without imposing SSR's, the mutual information and entropy of the mode reduced density operator \ref{eq:onemoderdospin} read:
\begin{equation} 
\label{eqn:2modecorrelation}
	\begin{aligned}
	I(\rho_{ \{m_\uparrow,m_\downarrow \}}) &= 2S(\rho_{ \{m_\uparrow,m_\downarrow \}})= -2\sum_{i}q_i \ln(q_i), \\
		E(\rho_{ \{m_\uparrow,m_\downarrow \}}) &= S(\rho_{ \{m_\uparrow,m_\downarrow \}}) = -\sum_{i}q_i \ln(q_i).
	\end{aligned}
\end{equation}

A detailed calculation presented in Ref.~\cite{Ding2020} establishes the density operators accessible under consideration of P-SSR and N-SSR.
 Under local parity constraints, the physical part of the density operator $\rho^P$ can be expressed as:
\begin{equation}
\begin{aligned}
\rho^{P} =\;& (q_1 + q_2)\ket{\Psi_{odd}} \bra{\Psi_{odd}} \\ 
    &+ (q_0 + q_3)\ket{\Psi_{even}} \bra{\Psi_{even}},\\
	\ket{\Psi_{odd}} =\;& \sqrt{\frac{q_1}{q_1+q_2}} \ket{m_\uparrow} \otimes \ket{\psi_{N-1}}  \\
    &+ \sqrt{\frac{q_2}{q_1+q_2}} \ket{m_\downarrow} \otimes \ket{\psi_{N-1}'}, \\
	\ket{\Psi_{even}} =\; & \sqrt{\frac{q_0}{q_0+q_3}} \ket{\Omega} \otimes \ket{\psi_{N}} \\
    &+ \sqrt{\frac{q_3}{q_0+q_3}} \ket{m_\uparrow, m_\downarrow} \otimes \ket{\psi_{N-2}}.
\end{aligned}
\end{equation}
Similarly, the introduction of local particle number constraints gives rise to the physically accessible density operator $\rho^N$:
	\begin{equation}
	\begin{aligned}
		\rho^{N} &=q_0\ket{\Psi_{0}} \bra{\Psi_{0}} + (q_1+q_2)\ket{\Psi_{1}} \bra{\Psi_{1}} + q_3\ket{\Psi_{2}} \bra{\Psi_{2}}, \\
	\ket{\Psi_{0}} &= \ket{\Omega} \otimes \ket{\psi_{N}}, \\ \ket{\Psi_{1}} &= \ket{\Psi_{odd}}, \\
	\ket{\Psi_{2}} &= \ket{m_\uparrow, m_\downarrow} \otimes \ket{\psi_{N-2}}.
	\end{aligned}
	\end{equation}
    Consequently, the physically accessible mutual information can be written as:
	\begin{equation}
	\label{eqn:2modePN-I}
	\begin{aligned}
		I^{PSSR}(\rho_{ \{m_\uparrow,m_\downarrow \}})= -2\sum_{i} q_i \ln(q_i) + (q_1 + q_2)\ln(q_1+q_2)\\ +(q_0 + q_3)\ln(q_0+q_3), \\
		I^{NSSR}(\rho_{ \{m_\uparrow,m_\downarrow \}})= -2\sum_{i} q_i \ln(q_i) + (q_1 + q_2)\ln(q_1+q_2) \\+q_0\ln(q_0) + q_3\ln(q_3).
	\end{aligned}
	\end{equation}
	The physically accessible entanglement between a spin-ful mode subsystem and the remaining system is given by:
	\begin{equation}
		\label{eqn:2modePN-E}
    	\begin{aligned} 
    	E^{PSSR}(\rho_{ \{m_\uparrow,m_\downarrow \}}) =& -\sum_{i} q_i \ln(q_i)+(q_1+q_2)\ln(q_1+q_2) \\ &+(q_0+q_3)\ln(q_0+q_3), \\
    	E^{NSSR}(\rho_{ \{m_\uparrow,m_\downarrow \}}) =& (q_1+q_2)\ln(q_1+q_2) - q_2\ln(q_2) \\ &-q_1\ln(q_1).
    	\end{aligned}
	\end{equation}
It shall be stressed, that only due to the spin-symmetries of this particular configuration it becomes possible to determine analytic expressions of the relative entropy of entanglement. In general, in order to find the closest separable state, a minimisation over a possibly large parameter space is necessary.

\section{Analytic expressions for the coefficients of the fermionic mode reduced density operator}
\label{sec:analyticqi}
 One can express the coefficients of the mode-reduced density operator in Eq.~\eqref{eq:SchmidtDecomposition} as a function of the dimensionless constant $\kappa$ (cf. Eq.~\eqref{eq:kappa}). It is illustrative to consider the three-fermion Harmonium ground state in one spatial dimension, as the expressions become increasingly complex for increasing particle numbers and dimensions. For mode $0$ the coefficients of the mode RDO of three-Harmonium read:

\begin{equation}
    \begin{aligned}
    q_1 &= \frac{2 \sqrt{a} \sqrt{\frac{a (a-3 b)}{a-2 b}}}{\sqrt{\frac{2 a^2-5 a b+b^2}{a-2 b}} \sqrt{\frac{a
\left(4 a^2-12 a b+5 b^2\right)}{2 a^2-5 a b+b^2}}}, \\
q_2 &= \frac{2 a^{3/2} \sqrt{\frac{a (a-3 b)}{a-2 b}} \left(4 a^2-12 a b+7 b^2\right)}{(a-2
b) \left(\frac{2 a^2-5 a b+b^2}{a-2 b}\right)^{3/2} \left(\frac{a \left(4 a^2-12 a b+5 b^2\right)}{2 a^2-5 a b+b^2}\right)^{3/2}},\\
    q_3 &=  \frac{\sqrt{a-3 b} (2 a-3 b)^2 \sqrt{\frac{4 a^2-6 a b+b^2}{a-b}} \sqrt{\frac{a \left(2
a^2-4 a b+b^2\right)}{4 a^2-6 a b+b^2}}} {4a(a-2 b)^2 (a-b)^{\frac{3}{2}}} \\
&\times \sqrt{\frac{a \left(a^2-3 a b+2 b^2\right)}{4 a^2-10 a b+5 b^2}} \sqrt{\frac{a \left(4 a^2-10 a b+5 b^2\right)}{2
a^2-4 a b+b^2}} ,\\
q_0&=1-q_1-q_2-q_3,
    \end{aligned}
\end{equation}
where $a=\frac{1}{2\tilde{l}^2}$ and $b = -\frac{1}{6} \left( \frac{1}{l^2}-\frac{1}{\tilde{l}^2} \right)$ and $N=3$ in this particular case. 

\section{Fermionic spatial mode entanglement: additional plots}\label{sec:AppSingleModeFerm}

\begin{figure}[h] 
\includegraphics[width=0.52\textwidth]{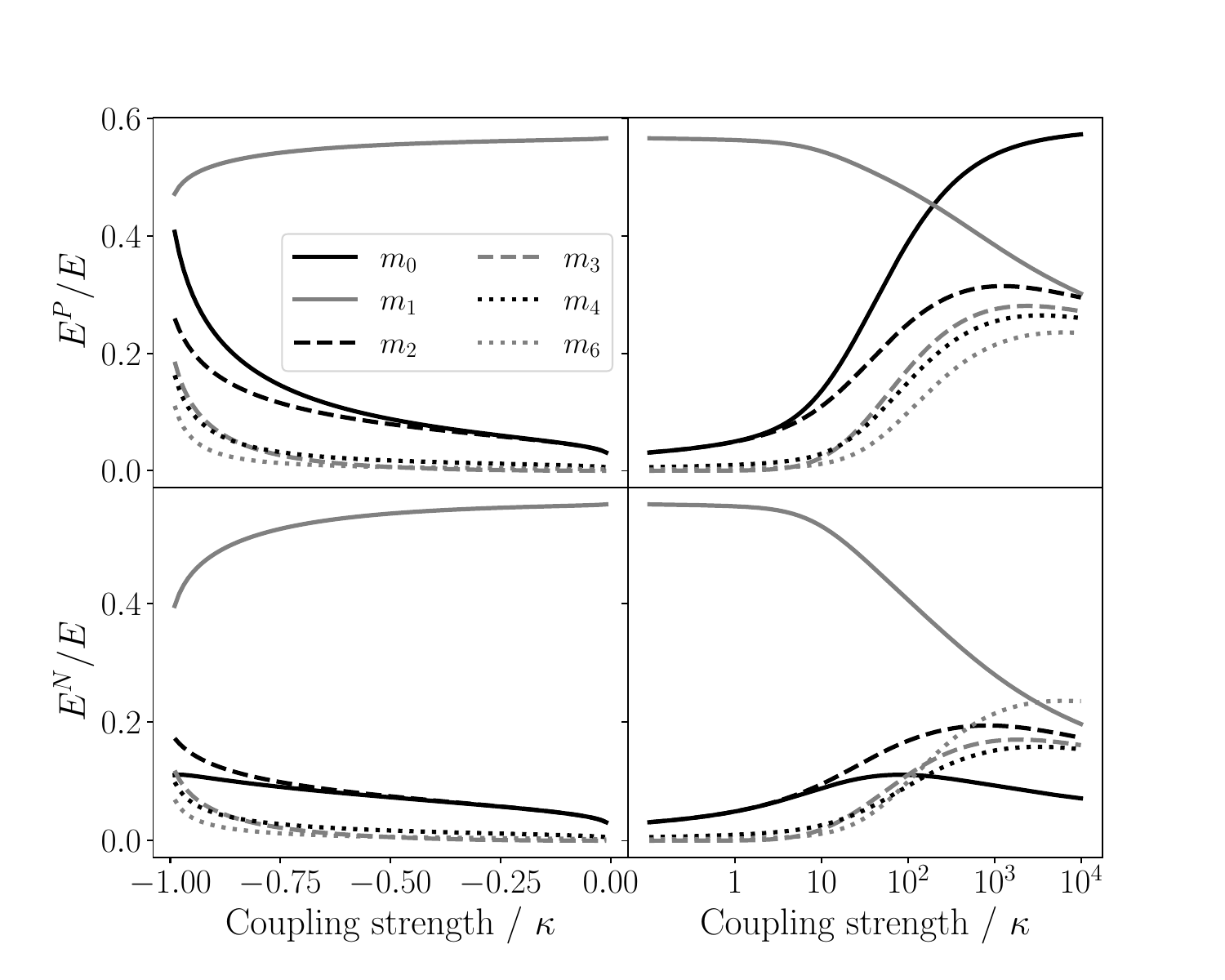}
\caption{Relative amount of physically accessible entanglement $E^{P/N}/E$ of spatial modes $\phi_0$ to $\phi_6$ as a function of the coupling strength $\kappa$ in one-dimensional 3-particle Harmonium.}
\label{fig:epnssr_fracn3}
\end{figure}

\begin{figure}[h] 
\includegraphics[width=0.52\textwidth]{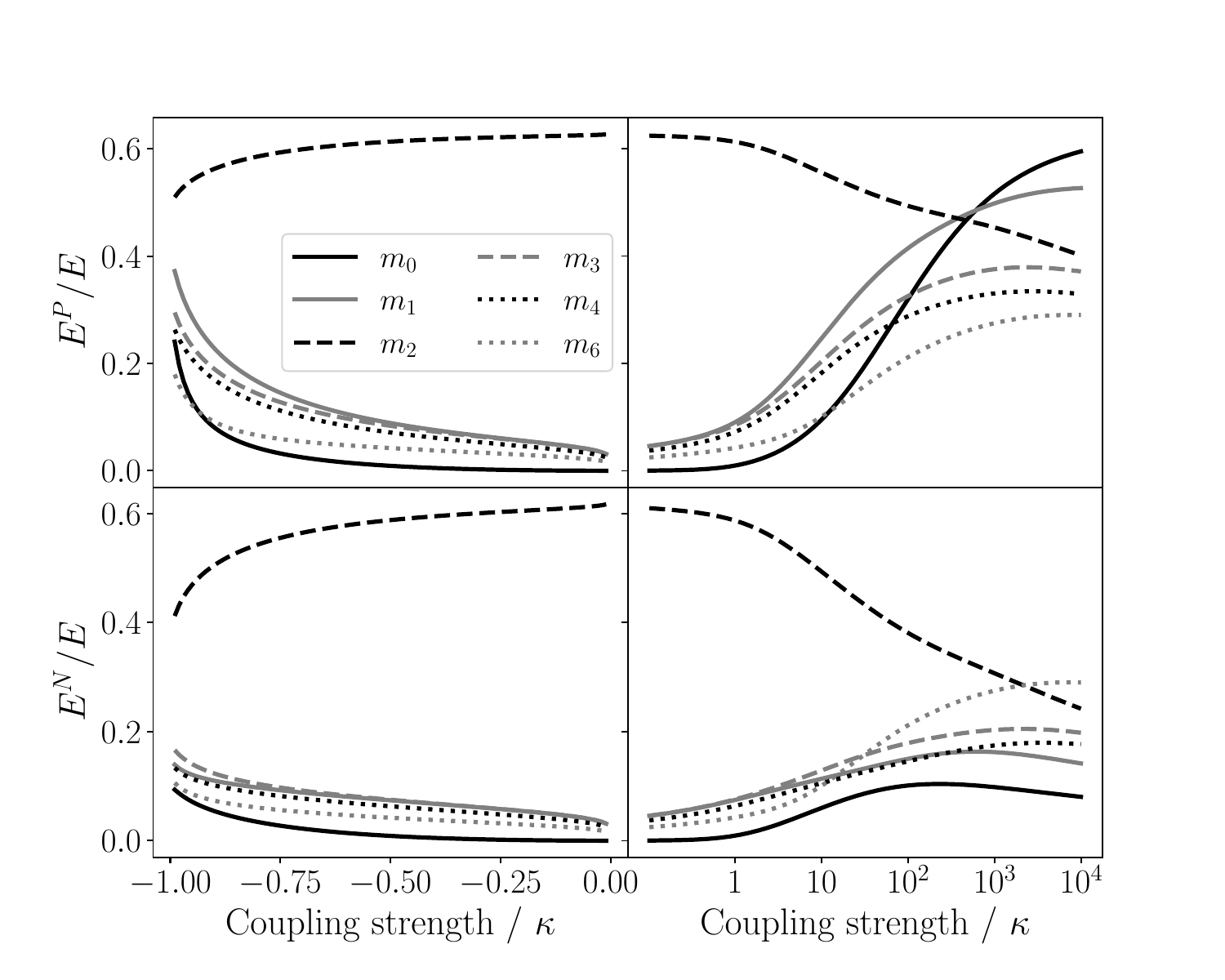}
\caption{Relative amount of physically accessible entanglement $E^{P/N}/E$ of spatial modes $\phi_0$ to $\phi_6$ as a function of the coupling strength $\kappa$ in one-dimensional 5-particle Harmonium.}
\label{fig:epnssr_fracn5}
\end{figure}

In this appendix, we display additional graphics on mode entanglement. These are Figs.~\ref{fig:epnssr_fracn3} and \ref{fig:epnssr_fracn5}. 

  \section{Fermionic mode-mode entanglement: deriving the two mode-reduced density operator and introducing local SSR's}
\label{sec:calc-f-mm-ent}
The fermionic two-mode reduced density operator is $16 \times 16$ dimensional and can be divided into $5$ block diagonal sectors of fixed global particle number $\lbrace N=0,1,2,3,4 \rbrace$. The elements of the two-mode reduced density operator can be written as expectation values of the creation and annihilation operators which allows for an explicit matrix representation. For the local basis $ \lbrace{ \ket{i_{\uparrow}}, \ket{i_{\downarrow}}, \ket{j_{\uparrow}}, \ket{j_{\downarrow}} \rbrace}$ with corresponding creation operators $\lbrace{ f_1^{\dagger}, f_2^{\dagger}, f_3^{\dagger}, f_4^{\dagger} \rbrace}$ this yields:

\begin{equation}
\label{eq:mmrdo_exp}
\begin{aligned}
(& \rho_{m_{\uparrow \downarrow}m_{\uparrow \downarrow}})_{n_1', n_2', n_3', n_4'}^{n_1, n_2, n_3, n_4} = \\
\langle &(f_1^{\dagger})^{n_1'} (f_2^{\dagger})^{n_2'} (f_3^{\dagger})^{n_3'} (f_4^{\dagger})^{n_4'} f_4 f_3 f_2 f_1 \cdot \\
&\cdot f_1^{\dagger} f_2^{\dagger} f_3^{\dagger} f_4^{\dagger}
(f_4^{\dagger})^{n_4} (f_3^{\dagger})^{n_3} (f_2^{\dagger})^{n_2} (f_1^{\dagger})^{n_1} \rangle
\end{aligned}
\end{equation}

The super-indices represent the occupation of various spin-ful modes. Non-vanishing expectation values only exist for blocks of fixed global particle number $N$ such that $n_1+n_2+n_3+n_4=n_1'+n_2'+n_3'+n_4'=N$. For example, the expectation value corresponding to the matrix element $\ket{i_{\uparrow, \downarrow}} \otimes \ket{\Omega} \bra{i_{\uparrow, \downarrow}} \otimes \bra{\Omega}$ would correspond to the indices $n_1=n_2=n_1'=n_2'=1, n_3=n_4=n_3'=n_4'=0$. Practically speaking, expanding the expression in Eq.~\eqref{eq:mmrdo_exp} gives rise to integrals of the various particle reduced density operators (cf. Eq.~\eqref{eq:part_red_dens_operator}), in analogy to the explicit expressions outlined in \ref{sec:expval_int}.

The introduction of a local number or parity constraint severely constrains the amount of entanglement between the modes. The physically accessible entanglement arises from particular sub-sectors of the two-mode reduced density operator and reads:
\begin{equation}
\label{eq:modemodeEN}
\begin{aligned}
     E(\rho_{m_{\uparrow \downarrow}m_{\uparrow \downarrow}}^{N}) &= S(\rho_{m_{\uparrow \downarrow}m_{\uparrow \downarrow}}^{N} \| \sigma_N ) \\
 E(\rho_{m_{\uparrow \downarrow}m_{\uparrow \downarrow}}^{P}) &= S(\rho_{m_{\uparrow \downarrow}m_{\uparrow \downarrow}}^{N} \| \sigma_N ) + S(\rho_{m_{\uparrow \downarrow}m_{\uparrow \downarrow}}^{even} \| \sigma_{even} ).
 \end{aligned}
 \end{equation}
 Here, the sector with fixed local particle number $\rho_{m_{\uparrow \downarrow}m_{\uparrow \downarrow}}^{N}=\rho_{m_{\uparrow \downarrow}m_{\uparrow \downarrow}}^{odd}$ coincides with the odd particle number sector and is spanned by: $\lbrace{\ket{i_\uparrow} \otimes \ket{j_\uparrow}, \ket{i_\uparrow} \otimes \ket{j_\downarrow}, \ket{i_\downarrow} \otimes \ket{j_\uparrow}, \ket{i_\downarrow} \otimes \ket{j_\downarrow} \rbrace}$. The sector with even local parity $\rho_{m_{\uparrow \downarrow}m_{\uparrow \downarrow}}^{even}$ is spanned by $\lbrace{\ket{\Omega} \otimes \ket{\Omega},  \ket{i_{\uparrow \downarrow}} \otimes \ket{\Omega}, \ket{\Omega} \otimes \ket{j_{\uparrow \downarrow}},  \ket{i_{\uparrow, \downarrow}} \otimes \ket{j_{\uparrow \downarrow}} \rbrace}$. The PPT-criterion here is a sufficient condition for separability \cite{ppt_separability} and thus, we use it to determine the closest separable states $\sigma_N$ and $\sigma_{even}$.

\end{document}